\PassOptionsToPackage{dvipsnames}{xcolor}
\PassOptionsToPackage{hyphens}{url}
\documentclass[sigconf, screen, noacm]{acmart}

\usepackage{xspace}
\usepackage{tikz}
\usepackage[ruled,lined]{algorithm2e}
\usepackage{booktabs}
\usepackage{soul}

\setcopyright{none}
\acmISBN{978-x-xxxx-xxxx-x/YY/MM}
\acmDOI{XXXXXXX.XXXXXXX}
\newcommand{\zhuhe}[1]{{\textcolor{blue}{\textit{#1}}}}

\newcommand{\orange}[1]{{\textcolor{Melon}{\textbf{#1}}}}
\newcommand{\blue}[1]{{\textcolor{SkyBlue}{\textbf{#1}}}}

\newcommand{\PBox}[1]{%
  \par
  \noindent\fbox{%
    \parbox{\dimexpr\linewidth-2\fboxsep-2\fboxrule\relax}{#1}%
  }%
  \par
}

\newcommand{\para}[1]{\textit{\textbf{#1}} }
\newcommand{\inlineheading}[1]{\underline{\textit{#1}} }

\newcommand{\Sect}[1]{Sec.~\ref{#1}}
\newcommand{\Fig}[1]{Fig.~\ref{#1}}
\newcommand{\Tbl}[1]{Tbl.~\ref{#1}}

\newcommand{\Eqn}[1]{Eqn.~\ref{#1}}

\newcommand{\Alg}[1]{Algo.~\ref{#1}}

\newcommand{\specialcell}[2][c]{\begin{tabular}[#1]{@{}c@{}}#2\end{tabular}}

\newcommand{\mode}[1]{\underline{\textsc{#1}}\xspace}

\newcommand{\proj}{\textsc{ElsaGS}\xspace}
\newcommand{\cache}{\textsc{FGcache}\xspace}

\definecolor{myorange}{RGB}{255, 212, 121}

\graphicspath{{figures/}}

\begin{document}

\title{Accelerating 3D Gaussian Splatting Training through Selective Freezing and Computation Caching}

\author{He Zhu}
\orcid{0009-0006-0835-7153}
\affiliation{%
  \institution{Shanghai Jiao Tong University}
  \city{Shanghai}
  \country{China}
}
\email{zhcon16@sjtu.edu.cn}

\author{Ziheng Dai}
\orcid{}
\affiliation{%
  \institution{Shanghai Jiao Tong University}
  \city{Shanghai}
  \country{China}
}
\email{zhihengdai@sjtu.edu.cn}

\author{Zihan Liu}
\orcid{}
\affiliation{%
  \institution{Shanghai Jiao Tong University}
  \city{Shanghai}
  \country{China}
}
\email{altair.liu@sjtu.edu.cn}

\author{Ruyang Li}
\orcid{}
\affiliation{%
  \institution{IEIT SYSTEMS Co., Ltd.}
  \city{Beijing}
  \country{China}
}
\email{liruyang@ieisystem.com}

\author{Hui Wei}
\orcid{}
\affiliation{%
  \institution{IEIT SYSTEMS Co., Ltd.}
  \city{Beijing}
  \country{China}
}
\email{weihui@ieisystem.com}

\author{Yaqian Zhao}
\orcid{}
\affiliation{%
  \institution{IEIT SYSTEMS Co., Ltd.}
  \city{Beijing}
  \country{China}
}
\email{zhaoyaqian@ieisystem.com}

\author{Han Zhao}
\orcid{}
\affiliation{%
  \institution{Shanghai Jiao Tong University}
  \city{Shanghai}
  \country{China}
}
\email{zhao-han@cs.sjtu.edu.cn}

\author{Xiaofeng Hou}
\orcid{}
\affiliation{%
  \institution{Shanghai Jiao Tong University}
  \city{Shanghai}
  \country{China}
}
\email{hou-xf@cs.sjtu.edu.cn}

\author{Chen Chen}
\orcid{}
\affiliation{%
  \institution{Shanghai Jiao Tong University}
  \city{Shanghai}
  \country{China}
}
\email{chen-chen@sjtu.edu.cn}

\author{Jingwen Leng}

\orcid{0000-0002-5660-5493}
\affiliation{%
  \institution{Shanghai Jiao Tong University}
  \city{Shanghai}
  \country{China}
}
\email{leng-jw@sjtu.edu.cn}

\author{Yu Feng}
\authornote{Corresponding Author.}
\orcid{0000-0002-2192-5737}
\affiliation{%
  \institution{Shanghai Jiao Tong University}
  \city{Shanghai}
  \country{China}
}
\email{y-feng@sjtu.edu.cn}

\author{Minyi Guo}
\orcid{0000-0003-0034-2302}
\affiliation{%
  \institution{Shanghai Jiao Tong University}
  \city{Shanghai}
  \country{China}
}
\email{guo-my@cs.sjtu.edu.cn}

\renewcommand{\shortauthors}{He Zhu et al.}

\begin{abstract}

3D Gaussian Splatting (3DGS) becomes a key primitive for immersive 3D experiences. 
However, its applicability is limited by its long training overhead.
Although prior work has shown that gradient accumulation is the dominant bottleneck in training, existing works either have limited speedup or degrade reconstruction quality.

To address this issue, we present \proj, an algorithm–hardware co-designed framework that accelerates 3DGS training with negligible accuracy degradation.
Our key observation is that most Gaussians converge early, while later iterations only refine a small subset of Gaussians. 
Thus, we propose a technique, \textit{progressive Gaussian freezing}, which dynamically freezes well-converged Gaussians and caches their cumulative contributions, thereby reducing computations in both forward rasterization and backward gradient accumulation.
To make caching efficient, we further propose an architecture, \proj core, which integrates a pixel-parallel rasterization array with a set-associative frozen-Gaussian cache, \cache.
Across three 3DGS datasets, \proj freezes up to 90\% of Gaussians with negligible quality loss and achieves up to 20.6$\times$ end-to-end speedup and 99.4\% energy reduction over GPUs.

\end{abstract}
\maketitle

\section{Introduction}
\label{sec:intro}

3D Gaussian Splatting (3DGS)~\cite{kerbl20233d, wu2024recent, fang2024mini, lee2023compact, fan2023lightgaussian} has emerged as a next-generation primitive for spatial computing and immersive 3D social experiences.
Recent VR platforms, e.g., Meta Quest 3, already integrate 3DGS with applications, e.g., Arrival Space~\cite{3dgs_app}, and enable users to relive scenes interactively.
Huawei and Kiri recently introduced Remy~\cite{remy} into HarmonyOS, allowing everyday users to reconstruct high-fidelity spatial content directly using their smartphones. 
These applications fundamentally change how users record and share their memories, transforming photos into interactive 3D experiences rather than flat images.

Despite its promising user experience, 3DGS is still limited by its training time.
E.g., reconstructing even a small-scale scene, \textit{Room}, requires around 60 minutes on a mobile Nvidia Ampere GPU~\cite{orinsoc}.
This means that users must wait tens of minutes before viewing or sharing captured scenes, breaking the immediacy expected from mobile and social applications. 
Therefore, accelerating the 3DGS training process is critical for interactive spatial memory capturing.

To accelerate 3DGS training, prior studies target backward gradient accumulation, where numerous pixels update the same Gaussian parameters through costly atomic operations~\cite{durvasula2025arc, he2025gsarch, huang2026splatonic, mallick2024taming, wen2026cambricon, wu2024gauspu}.
Existing solutions attempt to alleviate this overhead either by redesigning computational  dataflows~\cite{mallick2024taming, wen2026cambricon, huang2026splatonic, durvasula2025arc} or by pruning insignificant gradients~\cite{he2025gsarch, wu2024gauspu}.
However, those approaches typically have limited speedup or introduce noticeable quality degradation.

\para{Motivation.}
Our goal is to speed up training by orders of magnitude while preserving accuracy.
We notice that \textit{a large portion of Gaussians stabilize early with low gradients, while subsequent iterations predominantly refine a small active subset of Gaussians} (\Sect{sec:motiv}).
This creates an opportunity to stop their updates and reuse their per-view contributions across iterations to reduce computation in both forward and backward passes during training.

\para{Framework.} We thus propose \proj, a \textit{progressive Gaussian-freezing framework} in \Sect{sec:framework} that ``freezes'' low-gradient Gaussians dynamically while continuing to optimize the remaining active ones.
The framework combines two optimizations. First, it caches their per-view contributions to avoid repeated computation in both the forward and backward passes over iterations. Second, it skips backpropagation for frozen Gaussians to eliminate their gradient accumulations.
\Sect{sec:eval:acc} shows that \proj achieves better reconstruction quality than prior gradient pruning techniques~\cite{he2025gsarch}.

\begin{figure*}
    \centering
    \includegraphics[width=\textwidth]{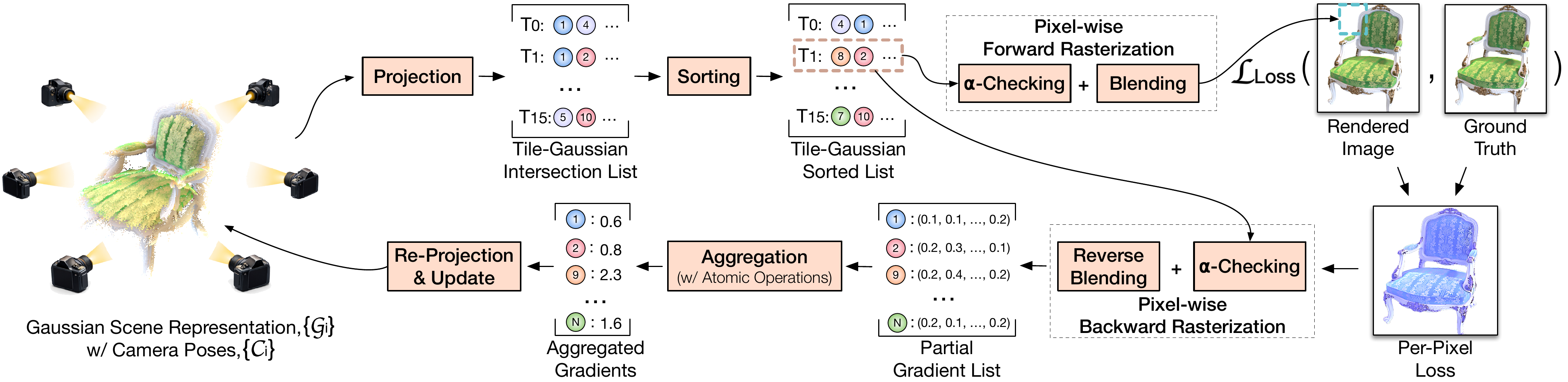}
    \caption{Overview of the 3DGS training pipeline, including the forward and backward passes. The inputs consist of a set of captured images and their corresponding camera poses obtained from photogrammetry tools. The optimization target is the Gaussian scene representation, i.e., a set of Gaussian primitives. In each training iteration, the forward pass renders an image from a selected camera pose and computes the pixel-wise loss against the ground-truth image. The backward pass then back-propagates this loss to individual Gaussians and updates their parameters accordingly.
    }
    \label{fig:training_overview}
\end{figure*}

\para{Caching.} 
While \proj avoids redundant computations, caching individual frozen-Gaussian contributions incurs substantial memory-access overhead. 
We therefore introduce \cache in \Sect{sec:framework:cache}, which stores cumulative contributions for consecutive frozen Gaussians.
To further make these group-level summaries usable by canonical backward rasterization, which only accepts individual contributions, we redesign gradient evaluation to consume the summaries directly.
Together, these designs reduce the data transmission overhead and enable cache reuse in both rasterization passes.


\para{Architecture.}
To support Gaussian lookups and updates, we propose a streaming architecture that pipelines different 3DGS stages while supporting sparse Gaussian updates in \Sect{sec:arch}.
The key innovation is the \proj core, which integrates the rasterization array with our dedicated cache, \cache, to enable efficient Gaussian reuse.
The rasterization array performs pixel-wise processing with on-chip partial gradient reduction to minimize off-chip memory traffic.
Meanwhile, \cache is designed to share the cached results across both forward and backward passes, further eliminating redundant storage overhead.
These optimizations allow \proj to reduce both rasterization and backpropagation overheads.

\para{Results.} 
We evaluate \proj on three widely used datasets~\cite{barron2022mipnerf360, Knapitsch2017, hedman2018deep}.
To demonstrate the effectiveness, we evaluate our approach on both canonical and pruned 3DGS algorithms~\cite{kerbl20233d, mallick2024taming}.
Our results show that \proj can freeze up to 90\% of Gaussians during training with negligible quality loss.
We further compare against off-the-shelf GPUs and dedicated 3DGS accelerators~\cite{he2025gsarch, wen2026cambricon}.
\proj delivers up to 20.6$\times$ speedup and 99.4\% energy savings.

The contributions of this paper are as follows:
\begin{itemize}
\item We propose \proj, a framework that accelerates training by progressive and selective Gaussian freezing.
\item We develop \cache and a cache-compatible backward rasterization to reuse contributions of frozen Gaussians.
\item We introduce an algorithm–hardware co-design that addresses the irregular memory access patterns introduced by sparse Gaussian updates via our cache architecture.
\item \proj achieves up to 20.6$\times$ speedup and 99.4\% energy savings against GPUs and prior accelerators.
\end{itemize}

\section{Background}
\label{sec:bg}


\subsection{3DGS Training Pipeline.}
\label{sec:bg:pipeline}

The goal of the 3DGS training process is to optimize the Gaussian primitives, $\{G_i\}$, that model the 3D scene.
Each Gaussian encodes scene attributes such as position, opacity, and color.
These Gaussian primitives are initially obtained by photogrammetry tools, e.g., COLMAP~\cite{schoenberger2016sfm}.
Then, the training process refines these Gaussians via iterations of the \textit{forward pass} and the \textit{backward pass} (\Fig{fig:training_overview}).

During the forward pass, the pipeline renders an image from a camera pose.
The backward pass then computes the loss between the rendered image and its corresponding ground-truth image, and back-propagates the loss to Gaussian parameters for parameter updates.
We next describe the key stages.

\para{Forward Pass.}
This pass contains \textit{projection}, \textit{sorting}, and \textit{forward rasterization}.
This pass renders images tile-by-tile.

\inlineheading{Projection.}
The projection stage filters out Gaussians that are outside the current view frustum.
In this tile-based pipeline, it determines the intersection between Gaussians and rendering tiles, and stores the results in a tile-Gaussian intersection list.

\inlineheading{Sorting.}
Gaussians within each tile are depth-sorted from nearest to farthest to ensure correct rendering order.

\inlineheading{Forward Rasterization.}
The sorted Gaussians are then rendered pixel-by-pixel.
Each pixel in a tile iterates over the Gaussians and performs $\alpha$-checking to determine whether a Gaussian contributes to that pixel.
A Gaussian is considered intersecting if its transparency $\alpha_i$ at the pixel exceeds a threshold $\alpha^*$.
The pixel color is then accumulated from its contributing Gaussians based on,
\begin{align}
\label{eqn:nerf}
   \mathbf{C}(\mathbf{p}) & = \sum_{i=1}^{N} \Gamma_i \alpha_i \mathbf{c}_{i},\ \text{and} \ \Gamma_i = \prod^{i-1}_{j=1} (1-\alpha_j),
\end{align}
where $\mathbf{C}(\mathbf{p})$ denotes the final color of pixel $\mathbf{p}$ and $\mathbf{c}_i$ represents the color of the $i$th Gaussian. $\alpha_i$ and $\Gamma_i$ are the transparency of the $i$th Gaussian and accumulated transmittance along the ray from the first Gaussian to the $(i-1)$th Gaussian at $\mathbf{p}$. 
The transparency $\alpha_i$ is determined by its opacity $\sigma_i$ and its 2D covariance matrix~\cite{zwicker2001ewa}.

\para{Backward Pass.}
Once the image is rendered, the backward pass first computes the pixel-wise loss, then backpropagates the loss to the Gaussians.
The backward pass consists of three stages: \textit{backward rasterization}, \textit{aggregation}, and \textit{re-projection}.

\inlineheading{Backward Rasterization.}
This stage computes the partial gradients of every pixel-Gaussian pair.
Similar to the forward pass, each pixel first identifies its intersecting Gaussians through $\alpha$-checking.
Given these per-pixel Gaussian lists, the backward rasterization inverts the color blending process (\Eqn{eqn:nerf}) and computes the partial gradients of individual Gaussians with respect to the pixel loss,
\begin{align}
\label{eqn:loss}
\frac{\partial L}{\partial \alpha_i}  (\mathbf{p}) & = {\nabla_{\mathbf{C}(\mathbf{p})}L}(\Gamma_i \mathbf{c}_{i} - \frac{1}{1-\alpha_i}\sum^{N}_{j = i+1}\alpha_j \mathbf{c}_j \Gamma_j),\\
\label{eqn:color_loss}
\frac{\partial L}{\partial \mathbf{c}_{i}} (\mathbf{p}) & = {\nabla_{\mathbf{C}(\mathbf{p})}L} \Gamma_i \alpha_i,
\end{align}
where $\frac{\partial L}{\partial \alpha_i}$ and $\frac{\partial L}{\partial \mathbf{c}_i}$ are the Gaussian-wise gradients for transparency and color, respectively, and ${\nabla_{\mathbf{C}(\textbf{p})}L}$ is pixel-wise color loss.
\textit{Backward rasterization processes Gaussians in reverse order} (from the $N$th Gaussian to the $1$st Gaussian) to propagate gradients.
Results are stored in the partial gradient list.

\begin{figure}
    \centering
    \includegraphics[width=\columnwidth]{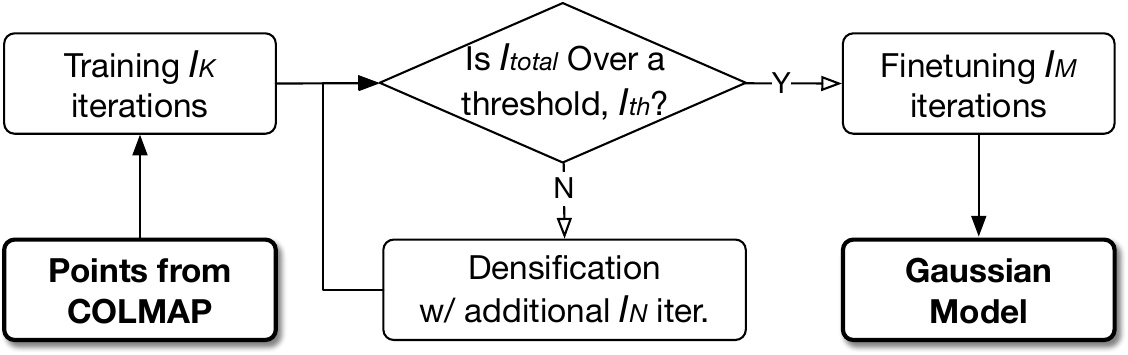}
    \caption{The general 3DGS training procedure. 
    It progressively optimizes Gaussians by iteratively performing training and periodic densification.
    Finally, a finetuning stage is performed to further refine rendering quality.}
    \Description{A flowchart loops between training and densification until the iteration threshold is reached. Finetuning then produces the final Gaussian model.}
    \label{fig:training_procedure}
\end{figure}

\inlineheading{Aggregation.}
The aggregation stage gathers all partial gradients associated with each Gaussian and accumulates them to obtain the final gradient for that Gaussian.
To ensure correct accumulation, existing GPU implementations use atomic operations.
This stage is often fused with backward rasterization.

\inlineheading{Re-Projection.}
Finally, the accumulated gradients are transformed from the camera coordinate system to the world coordinate system.
This computation is often lightweight.

\subsection{3DGS Training Procedure.}
\label{sec:bg:procedure}

With the training pipeline in \Sect{sec:bg:pipeline}, 3DGS adopts a progressive optimization procedure to produce photorealistic representations.
\Fig{fig:training_procedure} shows the workflow.
It begins with a 3D point cloud reconstructed by a photogrammetry pipeline, e.g., COLMAP, which is converted into initial Gaussian primitives.

\para{Procedure.}
Training proceeds iteratively.
The initial model is first optimized for $I_K$ iterations, during which Gaussian parameters are updated through repeated forward and backward passes.
After this stage, the algorithm checks whether the total number of training iterations, $I_{\text{total}}$, exceeds a threshold, $I_{\text{th}}$.

If the criterion is not satisfied, the system performs a densification step~\cite{kerbl20233d, mallick2024taming}.
During densification, the system clones or splits Gaussians with high gradients to better capture scene details, and removes those with low opacity.
After densification, the model continues training for an additional $I_N$ iterations to refine the scene.
This process repeats until the threshold, $I_\text{th}$, is reached.

Once the iteration threshold is satisfied, the training stops generating more new Gaussians and enters a final finetuning stage with $I_M$ training iterations.
This step refines Gaussian parameters to stabilize rendering quality until the training finishes.

\begin{figure}
    \centering
    \includegraphics[width=\columnwidth]{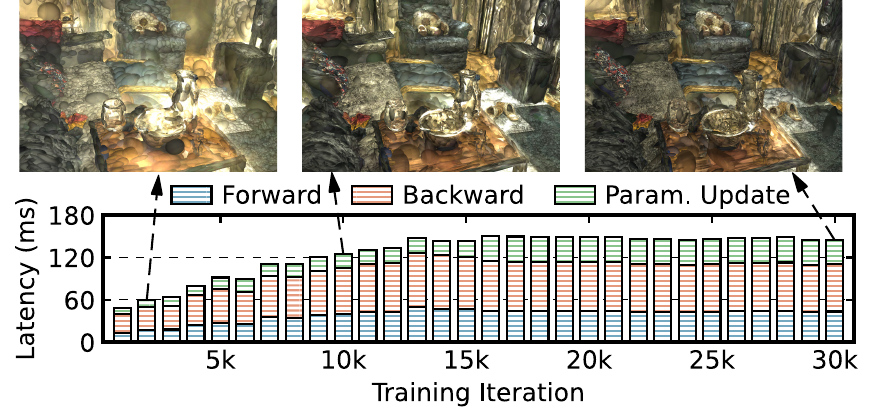}
    \caption{Profiled training-stage latency on \textit{Room} (Orin) with software atomic reduction~\cite{durvasula2025arc}. Top: visualization of Gaussian primitives at 2k, 10k, and 30k iterations. Bottom: stage latency averaged within 1k-iteration bins.}
    \Description{Three snapshots show Gaussian primitives at increasing training iterations. The latency traces below are plotted against training progress.}
    \label{fig:latency_profile}
\end{figure}

\para{Characteristics.}
During training, the initial stage $I_K$ is relatively short, 500-1000 iterations.
The following densification phase often consists of 10k-15k iterations, with each densification step occurring every 500 iterations.
Finally, the finetuning stage requires 15k-30k iterations.
For large-scale scenes, the finetuning can be even longer, up to 200k iterations~\cite{zhao2025on}.
\Fig{fig:latency_profile} shows the training-stage latency over iterations with atomic-reduction optimization~\cite{durvasula2025arc}.
Latencies of all stages steadily grow as training progresses due to the increasing number of Gaussians visualized by the three images in \Fig{fig:latency_profile}.


\section{Motivation}
\label{sec:motiv}

With an understanding of the 3DGS training, we next give two key observations that drive our algorithm-hardware co-design.


\begin{figure}
    \centering
    \includegraphics[width=\columnwidth]{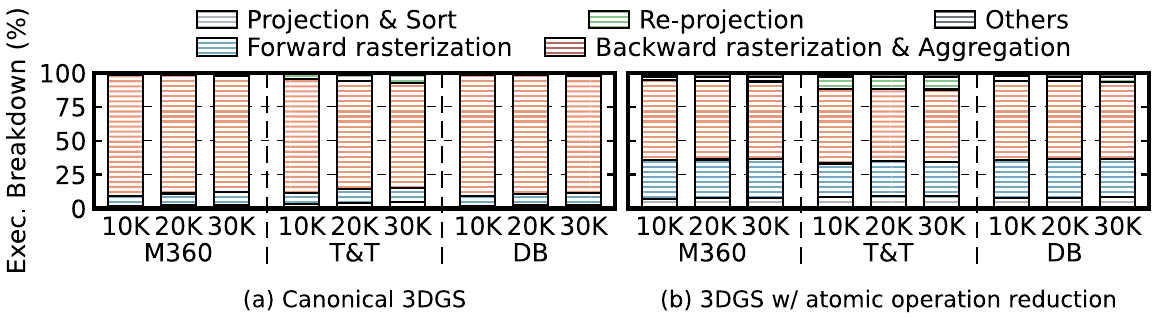}
    \caption{Execution breakdown of Canonical 3DGS ~\cite{kerbl20233d} and its variant with atomic operation reduction~\cite{durvasula2025arc} across three datasets over iterations 10k, 20k, and 30k.}
    \label{fig:exec_breakdown}
\end{figure}

\begin{figure}
    \centering
    \includegraphics[width=\columnwidth]{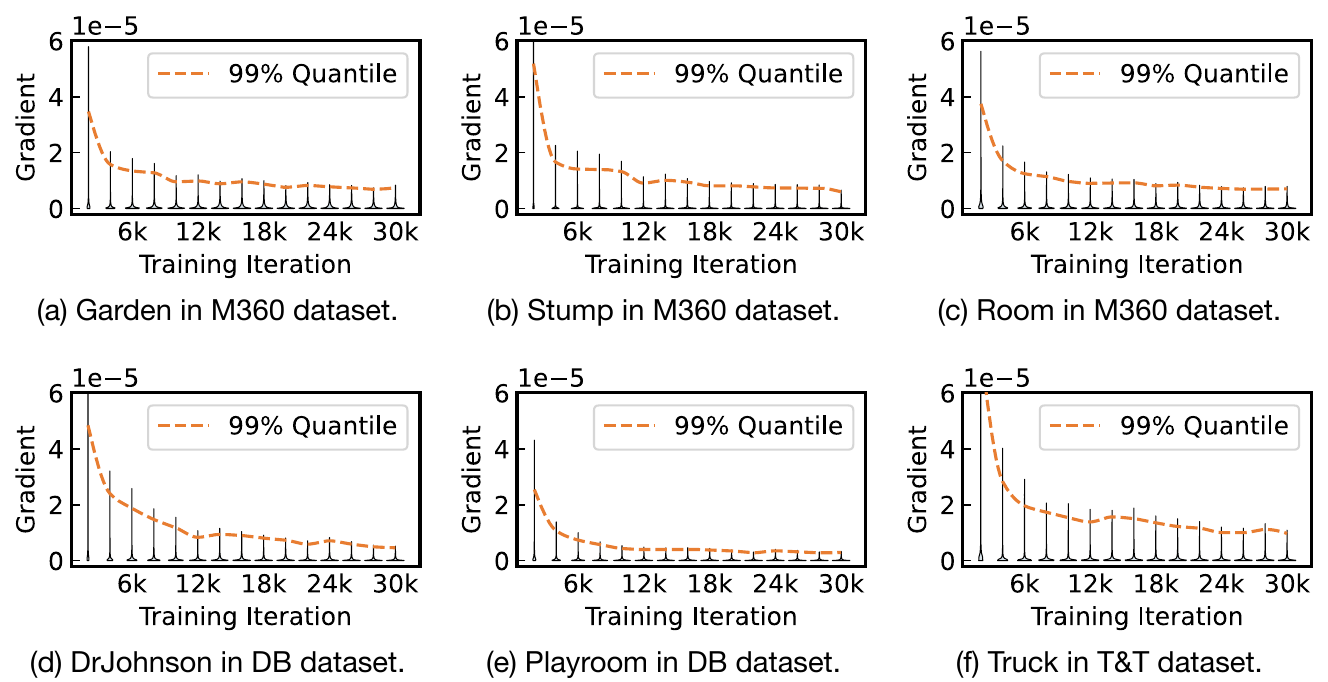}
    \caption{The absolute gradient distribution over the course of training. Here, we show six scenes across three datasets~\mbox{\cite{barron2022mipnerf360, hedman2018deep, Knapitsch2017}}. Most Gaussians have already converged after the initial training iterations.}
    \Description{Six scene panels compare gradient-magnitude distributions at different training iterations. The distributions concentrate near zero over time.}
    \label{fig:gradient_trend}
\end{figure}

\begin{figure*}
    \centering
    \includegraphics[width=\textwidth]{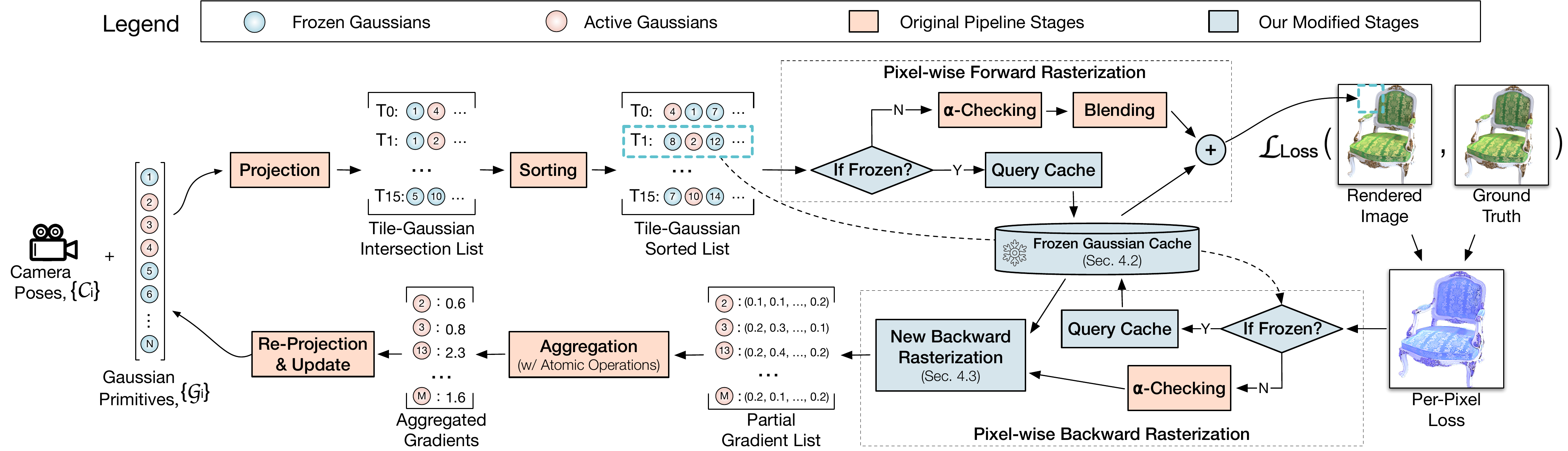}
    \caption{Our training pipeline with \textit{Gaussian freezing}. \orange{Orange} blocks denote stages in the original pipeline, while \blue{blue} blocks highlight our modifications. Orange and blue circles represent active and frozen Gaussians, respectively. Our training pipeline introduces two key modifications to reduce the overall computation. First, we cache the intermediate results of frozen Gaussians during forward rasterization. Second, we redesign the backward rasterization to enable cache sharing between forward and backward rasterization, allowing these two bottleneck stages to skip computation for grouped frozen Gaussians.
    }
    \Description{The forward and backward rasterizers branch on whether a Gaussian is frozen. The frozen paths connect to the same cache, and the active paths perform rasterization.}
    \label{fig:our_pipeline}
\end{figure*}

\PBox{\textbf{Observation 1:} Forward and backward rasterization dominate training, and the dominance persists as the model scales.}

\Fig{fig:exec_breakdown} shows the runtime breakdown of the forward and backward passes. Specifically, we profile three widely used datasets~\cite{barron2022mipnerf360, Knapitsch2017, hedman2018deep}, at 10k, 20k, and 30k training iterations on canonical 3DGS and variants with atomic operation reduction ~\cite{durvasula2025arc}.
Across all methods and training iterations, \textit{forward rasterization} and \textit{backward rasterization} dominate the total execution time.
Though method~\cite{durvasula2025arc} reduces atomic operations, the forward and backward rasterization are time-consuming, taking up over 80\% of the execution time. 
Both rasterization stages are computationally expensive since every pixel must traverse thousands of Gaussians during each iteration~\cite{feng2025lumina}.

\PBox{\textbf{Observation 2:} \textit{A majority of Gaussians converge initially during training, and only a small fraction of Gaussians require further training in subsequent iterations.}}

\mbox{\Fig{fig:gradient_trend}} presents six violin plots of the gradient magnitude distribution across all Gaussians during training.
The results are measured on three datasets~\cite{barron2022mipnerf360, hedman2018deep, Knapitsch2017}.
To reduce data noise, gradients are averaged over 100 iterations, and we plot every 2k iterations.
We consider a Gaussian to be well-trained once its gradient magnitude stabilizes and no longer decreases.

As shown in \Fig{fig:gradient_trend}, the distributions shift toward low magnitudes within the first 6k iterations, while their upper tails continue to decrease later in training.
This trend yields two key observations.
First, the majority of Gaussians converge early in training and no longer require further training after the initial training stage.
Second, a small subset of Gaussians still requires additional optimization to refine reconstruction details.
Meanwhile, the densification process would introduce new Gaussians that require further optimization.

Existing methods exploit early Gaussian convergence but either leave forward passes to be recomputed~\cite{armagan2026gsta} or rely on order-independent rendering for contribution reuse~\cite{yang2026sparseoit}.
In contrast, \proj reuses frozen Gaussian contributions in both forward and backward passes while preserving depth-ordered alpha blending.

\section{Progressive Gaussian-Freezing Framework}
\label{sec:framework}

With these two key observations in \Sect{sec:motiv}, we present our training framework that ``freezes'' the majority of Gaussians during training, i.e., stops updating their parameters.
We first give an overview of the training workflow in \Sect{sec:framework:ov}, and then describe our cache design that enables efficient execution in \Sect{sec:framework:cache}.
Lastly, we explain how we modify backward rasterization in \Sect{sec:framework:raster} to enable sharing a single cache between forward and backward passes.

\subsection{Overview}
\label{sec:framework:ov}

We first introduce our training pipeline that leverages frozen Gaussians to reduce computations, and then describe how these Gaussians are identified dynamically during training.

\para{Pipeline.} 
\Fig{fig:our_pipeline} shows our pipeline.
The orange blocks denote the stages in the original pipeline, while the blue blocks highlight our augmentations.
Specifically, our pipeline accelerates two bottlenecks, forward and backward rasterization.

During forward rasterization, the original pipeline computes the transparency of every Gaussian and blends its color contribution into individual pixels as defined in \Eqn{eqn:nerf}.
In contrast, our pipeline first determines whether a Gaussian is frozen, i.e., its parameters are held fixed and no longer updated.
For frozen Gaussians, we avoid their computations and directly query the \textit{Frozen Gaussian Cache}, \cache, to retrieve their precomputed contributions, e.g., transparency $\alpha_i$ and color $\mathbf{c}_i$.
For active Gaussians, the pipeline falls back to the canonical rasterization.
To further reduce caching overhead, we group consecutive frozen Gaussians and cache their cumulative contributions rather than individual Gaussian contributions.
\Sect{sec:framework:cache} further describes the detailed cache design.

Similarly, during backward rasterization, the pipeline needs to compute partial gradients of transparency $\alpha_i$ and color $\mathbf{c}_i$ (\Eqn{eqn:loss} and \Eqn{eqn:color_loss}).
In particular, canonical backward pipeline requires traversing the tile-Gaussian sorted list in reverse order and reconstructing the cumulative color and transmittance contribution through iteration of per Gaussian.
Here, we avoid the computation of frozen Gaussians by first checking whether a Gaussian is frozen.
If not, we directly compute the gradients for the active Gaussians; otherwise, we query \cache.
Note that, \cache stores cumulative contributions of consecutive frozen Gaussians rather than individual Gaussian contributions.
We thus redesign backward rasterization in \Sect{sec:framework:raster} to enable \cache sharing for backward rasterization.

\begin{figure}
    \centering
    \includegraphics[width=\columnwidth]{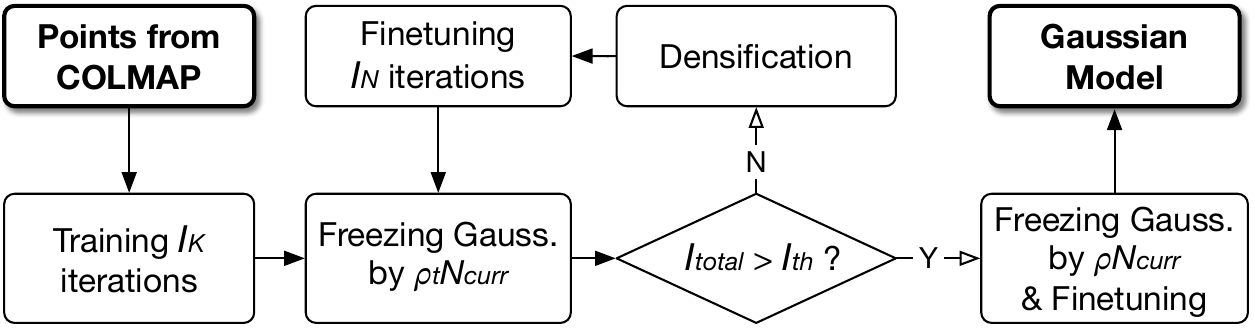}
    \caption{Our 3DGS training procedure. Before the finetuning stage, we iteratively freeze low-gradient Gaussians for every $I_{N}$ iterations, similar to the original training procedure. In the finetuning stage, we freeze $\rho$ of the Gaussians permanently until the total iteration threshold is reached.}
    \Description{The flowchart shows initial training, a loop that adds frozen Gaussians before densification, and a final finetuning path with a recovery interval.}
    \label{fig:our_procedure}
\end{figure}

\para{Procedure.}
\Fig{fig:our_procedure} shows the overall training procedure.
It consists of an \textit{initial training}, \textit{progressive freezing} during densification, and \textit{final finetuning}.
For the first 6k iterations, we follow the original training and densification schedule. After this, our framework begins to gradually freeze Gaussians.

In each densification iteration, we select low-gradient Gaussian candidates for freezing after removing low-opacity ones.
Similar to the densification process, we rank Gaussians using the same gradient statistic in densification and freeze those with the smallest scores.
All attributes of a frozen Gaussian are excluded from further optimization.
The freezing ratio $\rho_t$ is based on:
\begin{align}
\label{eqn:frozen}
    \rho_t &= \min\left(\rho,\kappa \max\left(\frac{N_\text{curr}-N_\text{init}}{N_\text{curr}}, 0\right)\right)
\end{align}
where $N_\text{init}$ is the initial COLMAP Gaussian count and $N_\text{curr}$ is the current count before densification.
$\rho$ caps the target freezing ratio, while $\kappa$ controls how quickly it increases with the Gaussian population. 
Before each densification, we compute a target frozen fraction $\rho_t$ using \Eqn{eqn:frozen}. 
This target applies to the total frozen population, including Gaussians frozen at earlier events. 
If the current frozen count is below the target, we freeze additional Gaussians.
Afterwards, we perform densification and initialize all newly added Gaussians as active, preventing them from being frozen immediately.
Next, we continue to train the remaining active Gaussians until the next densification iteration or the finetuning stage.

When densification ends at $I_\text{th}$, we enter the $I_M$-iteration finetuning stage.
At the start of the stage, we perform a final freezing step to bring the total frozen fraction to $\rho$. We then keep the resulting frozen set fixed for the remainder of training.
Lastly, training runs for an additional $I_M$ iterations and outputs the final Gaussian model.

In practice, we set $\rho=0.9$ and $\kappa=2.0$ to freeze up to 90\% of Gaussians, and any Gaussians frozen once will not be unfrozen.
Our evaluation in \Sect{sec:eval:acc} demonstrates that the procedure can preserve model quality without additional recovery of frozen Gaussians under our progressive freezing framework.

\subsection{Frozen Gaussian Cache, \cache}
\label{sec:framework:cache}

\begin{figure}
    \centering
    \includegraphics[width=\columnwidth]{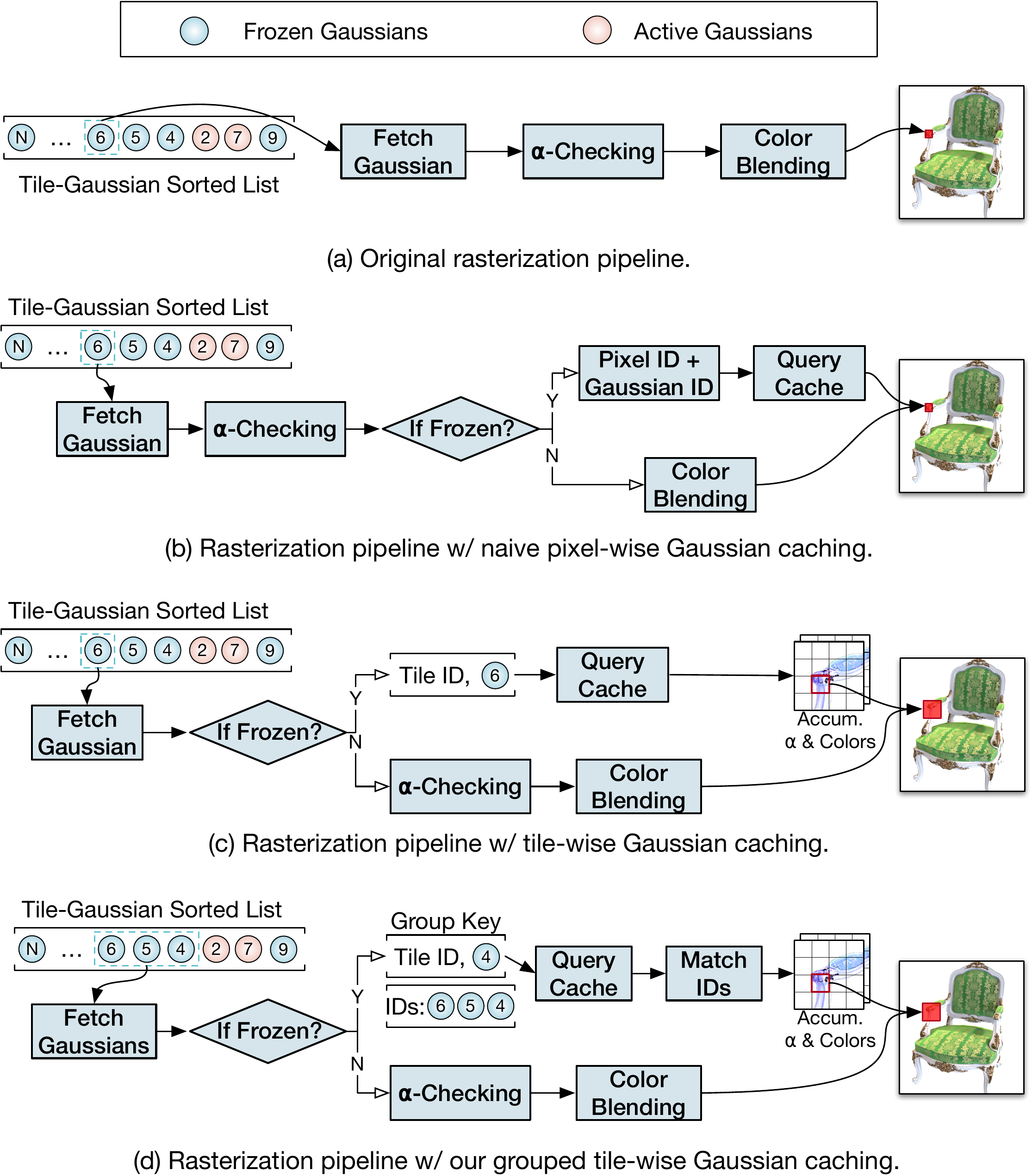}
    \caption{Intuition of \cache. Compared to naive Gaussian caching, i.e., storing every $\langle P_\text{ID}, G_\text{ID}\rangle$ pair, we introduce two optimizations: 1) grouping pixels and consecutive Gaussians to reduce metadata overhead, and 2) replacing $\langle T_\text{ID}, \{G_\text{ID}\}\rangle$ with a fixed-length $\langle T_\text{ID}, G_\text{ID, L} \rangle$ key. $P_\text{ID}$: pixel ID; $T_\text{ID}$: tile ID; $G_\text{ID}$: Gaussian ID; $G_\text{ID, L}$: leading Gaussian ID.}
    \Description{Four panels trace a sorted tile list through original rasterization, caching by pixel, caching by tile, and caching cumulative contributions for groups.}
    \label{fig:cache_idea}
\end{figure}

In this subsection, we introduce \cache.
Here, we describe the general caching abstraction.
The detailed architecture of \cache is shown in \Sect{sec:arch}.
We first present the key idea behind \cache and describe how intermediate results are cached, then describe our cache management policy.

\para{Idea.}
Here, \Fig{fig:cache_idea} illustrates the \cache design using forward rasterization of a single image as an example. 
The same caching principle applies to backward rasterization. 
\Fig{fig:cache_idea}a shows the canonical rasterization, where each pixel in a tile fetches individual Gaussians sequentially and blends their color contributions into the pixel.

A naive Gaussian caching approach would cache the intermediate results of individual Gaussians, including their transparencies $\alpha_i$ and color contributions $\mathbf{c}_i$, as shown in \Fig{fig:cache_idea}b. If a frozen Gaussian passes the $\alpha$-checking, its contribution to a pixel can be cached using the key $\langle P_\text{ID}, G_\text{ID}\rangle$, where $P_\text{ID}$ and $G_\text{ID}$ are the pixel ID and Gaussian ID. We then use this key to retrieve $\alpha_i$ and $\mathbf{c}_i$. However, caching every frozen Gaussian-pixel pair generates a large number of cache entries, leading to excessive memory traffic.

To reduce the number of cache entries, we organize these pixel-specific opacity, $\alpha_i$, and color contributions, $\mathbf{c}_i$, into one $\langle T_\text{ID}, G_\text{ID}\rangle$ entry per tile-Gaussian pair as shown in \Fig{fig:cache_idea}c, where $T_\text{ID}$ is the tile ID. 
Grouping these pixels reduces substantial repeated metadata reads and writes.
Nevertheless, tile-wise caching still requires a separate access for every frozen Gaussian.

To further reduce these accesses, we group consecutive frozen Gaussians by caching their cumulative contributions. 
\Fig{fig:cache_idea}d shows an instance where Gaussians 4, 5, and 6 form a consecutive frozen group.
We store the group's contribution under the key $\langle T_\text{ID}, \{G_\text{ID}\}\rangle$, where $\{G_\text{ID}\}$ is the set of consecutive frozen Gaussian IDs.
This design requires only one cache access per group.
However, explicitly recording all Gaussian IDs produces variable-length cache keys.

Fortunately, within a given tile's sorted Gaussian list, each consecutive group can be uniquely identified by its leading Gaussian ID, $G_\text{ID, L}$, and group length, since each Gaussian appears only once in the list.
We therefore use a fixed-length key, $\langle T_\text{ID}, G_\text{ID, L}\rangle$, to locate the cache entry and compare its stored group length with the current group length to confirm a group match.
In the cache policy below, we also show that this check remains effective when new updated active Gaussians change the Gaussian list.

\begin{figure}
    \centering
    \includegraphics[width=\columnwidth]{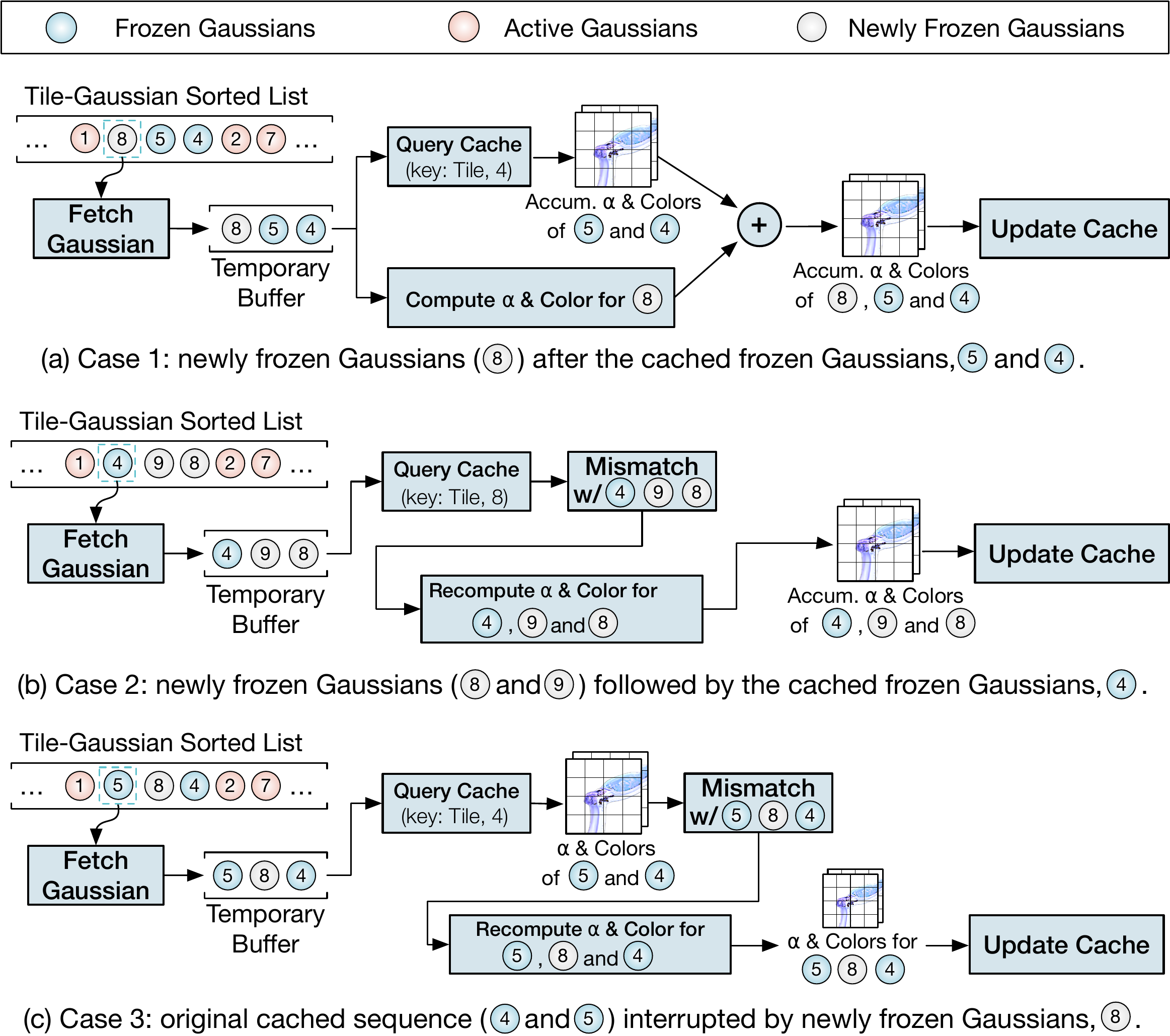}
    \caption{Three cache-update scenarios in \cache. An update is required when newly frozen Gaussians or active Gaussians change the rendering order of the sorted Gaussian list.}
    \Description{Three panels show changes in the order of Gaussians numbered 4, 5, and 8, with arrows from the temporary buffer to cache queries and contribution computations.}
    \label{fig:cache_update}
\end{figure}

\para{Cache Policy.}
In the following paragraphs, we describe the cache management policy of \cache.
Specifically, we explain four operations: \textit{read}, \textit{update}, \textit{insert}, and \textit{evict}.

\inlineheading{Read.}
We define a \textit{read} operation as querying results from \cache.
During rasterization, \cache maintains a temporary queue buffer filled with Gaussians from the sorted Gaussian list.
The rasterizer processes either a single unfrozen Gaussian or a consecutive frozen group at the top of the queue.
It queries \cache when the frozen group reaches the maximum group length, is followed by an unfrozen Gaussian, or reaches the end of the list.
Then, each consecutive frozen group is queried using $\langle T_\text{ID}, G_\text{ID, L}\rangle$.
Next, the length of the frozen Gaussians in the current temporary buffer is checked against the metadata stored in the cache entry.
A match means a cache hit.
Thus, we can directly reuse the cached result.
On a cache miss, we recompute all the Gaussians in the current temporary buffer, and follow the \textit{update} step described below to update the new cumulative contribution to \cache.

\inlineheading{Update.}
However, during training, the positions of the sorted Gaussian list and frozen groups can be changed by the updates to active Gaussians and newly frozen Gaussians. 
Cache entries must therefore be updated accordingly.
\Fig{fig:cache_update} shows three scenarios in which cache entries need to be updated.
In these examples, we assume the maximum frozen group length is 3.
For simplicity, some operations in the original rasterization are omitted.


\Fig{fig:cache_update}a shows the first case, where a newly frozen Gaussian, 8, follows the unchanged cached prefix of Gaussians 4 and 5.
According to the policy, we retrieve the entry using the key $\langle T_\text{ID}, 4\rangle$ and find its stored length of 2 is shorter than the current group length of 3.
Since the cached group remains an unchanged prefix of the current group, we reuse its cumulative contribution.
Meanwhile, the opacity and color contribution of Gaussian 8 are computed on the fly.
Lastly, we combine the computed contribution with the cached prefix in rendering order, and update the entry in place with the combined result and the new frozen group length of 3.

\Fig{fig:cache_update}b shows the second scenario, where an existing frozen Gaussian 4 is led by two newly frozen Gaussians, 8 and 9.
Since Gaussian 8 is newly frozen and not stored in cache, querying the new key $\langle T_\text{ID}, 8\rangle$ finds no valid entry, causing a cache miss.
We therefore recompute all three Gaussians and insert their cumulative contribution with a group length of 3.
The old subgroup starting at Gaussian 4 does not need explicit invalidation.
Since it is no longer accessed, it will eventually be evicted after rasterization.


In last scenario (\Fig{fig:cache_update}c), a frozen Gaussian, 8, moves between frozen Gaussians 4 and 5 and splits their cached group.
The current frozen group starting at Gaussian 4 ends before Gaussian 8, so querying $\langle T_\text{ID}, 4\rangle$ finds an entry with a longer stored length.
Thus, we recompute the contributions of all frozen Gaussians in the temporary buffer and update the cache with new results.

\inlineheading{Insert\&Eviction.}
In \cache, insertion occurs when new cumulative results need to be stored.
Unlike conventional caches that rely on least recently used (LRU) or similar replacement policies, we propose a new \textit{shortest Gaussian length} (SGL) replacement policy.
The policy is based on two observations.
First, during per-tile rasterization, the pipeline queries each frozen group exactly once. Thus, the temporary locality across entries is similar, and the primary difference is the amount of computation each entry can save.
Second, an entry that is not queried indicates that its cached group no longer corresponds to any current query group.
These entries are expired and should be removed to keep consistency.

Therefore, after each rasterization, any entries not queried are invalidated for that view and tile.
Upon insertion, \cache will first replace any invalid cache entry.
If \cache is full, SGL policy selects the entry with the smallest group of Gaussians for eviction.
This way, it retains entries that yield higher computational savings.

\begin{figure}
    \centering
    \includegraphics[width=\columnwidth]{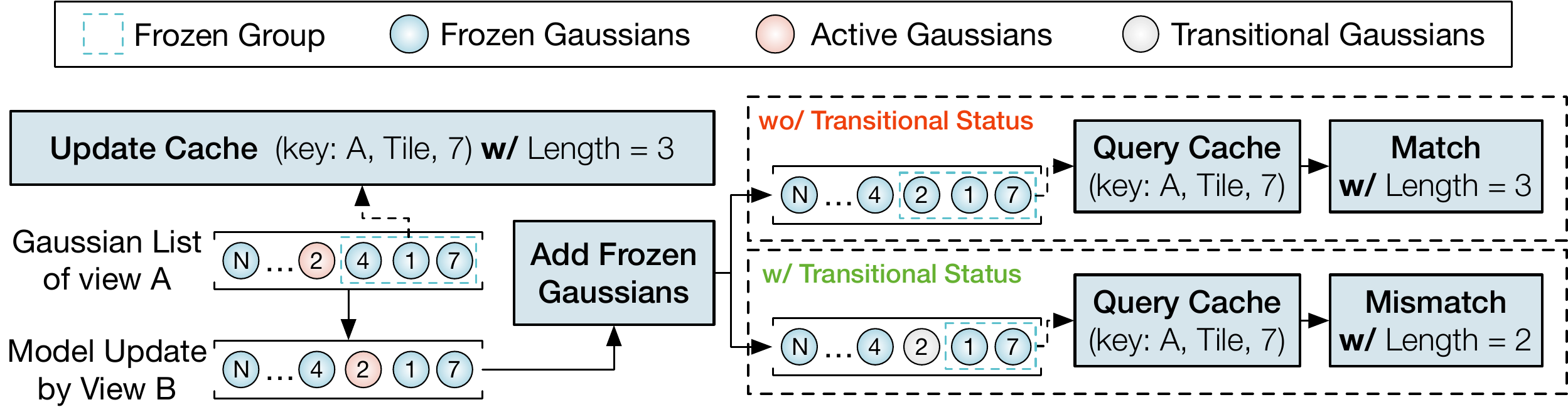}
    \caption{Cache refresh during multi-view training. An update from another view moves Gaussian 2 into a cached frozen group. With a maximum group length of 3, freezing Gaussian 2 immediately could change the group members while preserving the key and length. Here, we set Gaussian 2 as a transitional Gaussian, so the next cache lookup for this tile detects a mismatch between the current length of 2 and the cached length of 3 and triggers recomputation.}
    \Description{.}
    \label{fig:multiview_cache}
\end{figure}

\para{Multi-view Training.}
To support multi-view training, we add the image ID, $I_\text{ID}$, to form the key $\langle I_\text{ID}, T_\text{ID}, G_\text{ID, L}\rangle$, giving each view an independent cache space.
However, Gaussian updates across views can allow incorrect reuse of stale entries.
For a given view and tile, a cached frozen group may be split by an active Gaussian updated by another view, as shown in \Fig{fig:multiview_cache}.
This Gaussian may then be frozen before the corresponding entry is updated.
Since group lengths are capped, the current group may contain different members despite matching the leading ID and length.

To address this, we add one more training state, \textit{transitional}, to each Gaussian.
At each freezing step, selected active Gaussians enter the transitional state for one full freezing interval before promotion to frozen.
Transitional Gaussians have fixed parameters but remain unfrozen to the cache. Their contributions are computed individually rather than cached.
Their fixed positions preserve their relative order with existing frozen Gaussians throughout the interval.
To refresh every view's cache under the preceding policy before promotion, each freezing interval must cover all training views.
Our existing training configuration already satisfies this requirement, without changing the freezing schedule.

\subsection{New Backward Rasterization}
\label{sec:framework:raster}

\begin{algorithm}[t]
\caption{Our new backward rasterization.}
\label{algo:raster}
\KwData{Depth-ordered frozen-group and individual-Gaussian data for pixel $\mathbf{p}$, rendered color $\mathbf{C}(\mathbf{p})$, pixel color gradient $\nabla_{\mathbf{C}(\mathbf{p})}L$.}
$\mathbf{S}(\mathbf{p}) \leftarrow (0, 0, 0)$, $\Gamma \leftarrow 1$\;
\ForEach{item in forward order}{
    \eIf{item is a frozen group with data $(T_\text{group}, \mathbf{C}_\text{group})$}{
        $ \mathbf{S}(\mathbf{p}) \leftarrow \mathbf{S}(\mathbf{p}) + \Gamma \cdot \mathbf{C}_\text{group} $\;
        $ \Gamma \leftarrow \Gamma \cdot T_\text{group} $\;
    }{
        Let $G_i$ be the individual Gaussian with data $(\alpha_i, \mathbf{c}_i)$\;
        $ \mathbf{S}(\mathbf{p}) \leftarrow \mathbf{S}(\mathbf{p}) + \alpha_i \cdot \Gamma \cdot \mathbf{c}_i $\;
        \If{$G_i$ is trainable}{
            $ \frac{\partial L}{\partial \alpha_i}(\mathbf{p}), \ \frac{\partial L}{\partial \mathbf{c}_i}(\mathbf{p}) \leftarrow \text{ComputeGradients(}\nabla_{\mathbf{C}(\mathbf{p})}L\text{)}$\;
            $\text{AggregateGradients(}\frac{\partial L}{\partial \alpha_i}(\mathbf{p}), \frac{\partial L}{\partial \mathbf{c}_i}(\mathbf{p})\text{)}$\;
        }
        $ \Gamma \leftarrow \Gamma \cdot (1 - \alpha_i) $\;
    }
}
\end{algorithm}

While frozen Gaussians require no parameter gradients, their contributions are necessary for evaluating active-Gaussian gradients.
As \Eqn{eqn:loss} shows, the gradient $\frac{\partial L}{\partial \alpha_i}$ of an active Gaussian depends on the cumulative contributions of all Gaussians behind it. 
Canonical backward rasterization maintains this contribution through a back-to-front traversal, updating its running state using each Gaussian’s $\alpha$ and $\mathbf{c}$.
Frozen Gaussians must participate in these state updates, although their own gradient computations are omitted.
However, \cache only stores front-to-back group-level transmittance and color, without values required by the per-Gaussian updates.
To reuse the values cached in \cache, we reformulate gradient computation in terms of cumulative rendering contributions and design a backward rasterization that leverages cached frozen Gaussians.

\begin{figure*}
    \centering
    \includegraphics[width=\textwidth]{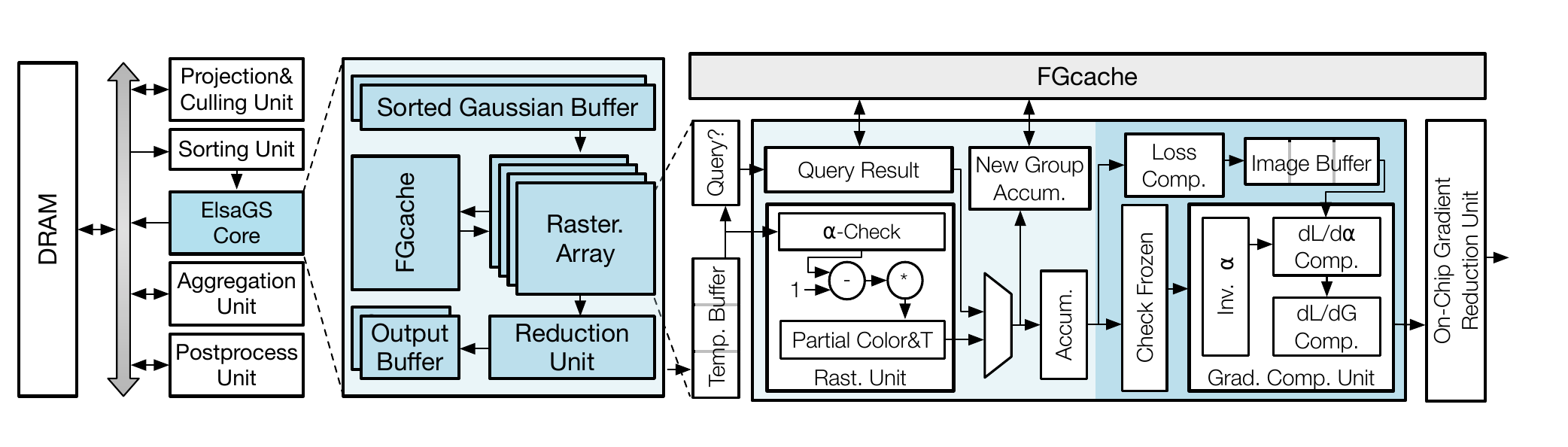}
    \caption{Overall streaming architecture of \proj. Our design integrates projection, sorting, aggregation, post-processing, and the \proj core into a pipelined dataflow. 
    Our key contribution is the \proj core, which has a set of rasterization units to accelerate both forward and backward rasterization, and a \cache to reuse the cumulative results of frozen Gaussians.}
    \Description{Projection, sorting, rasterization, and aggregation are connected in a pipeline. The core detail shows shared buffers and the cache feeding parallel rasterization units.}
    \label{fig:arch}
\end{figure*}

\para{Cache-Compatible Gradient Evaluation.}
In \Eqn{eqn:loss}, the term, $\sum^{N}_{j = i+1}\alpha_j \mathbf{c}_j \Gamma_j$, stands for the cumulative color contribution behind Gaussian ${i}$.
This quantity can be computed as the residual between the final pixel color and the forward accumulated color. 
Therefore, $\frac{\partial L}{\partial \alpha_i}$ can be refactored as,
\begin{equation}
    \frac{\partial L}{\partial \alpha_i} (\mathbf{p}) = {\nabla_{\mathbf{C}(\mathbf{p})}L} (\Gamma_i  \mathbf{c}_i - \frac{\mathbf{C}(\mathbf{p}) - \mathbf{S}_i(\mathbf{p})}{1 - \alpha_i} ),\ \mathbf{S}_i(\mathbf{p}) = \sum^{i}_{j=1}\alpha_j  \mathbf{c}_j  \Gamma_j.
\end{equation}
This formulation evaluates gradients using the running color and transmittance, which can be updated directly from \cache without reconstructing individual frozen-Gaussian contributions.

\para{Cache-Aware Backward Rasterization.}
We show our backward rasterization with Gaussian caching in \Alg{algo:raster}.
It implements the gradient evaluation through a front-to-back traversal and maintains a running color $\mathbf{S}(\mathbf{p})$ and a transmittance $\Gamma$.
Similarly, we follow the cache policy in \Sect{sec:framework:cache} to group consecutive frozen Gaussians and query their accumulated values from \cache.

For an active individual Gaussian, we compute and aggregate the gradient.
For a frozen group of $k$ Gaussians in \cache, we fetch the transmittance $T_\text{group}$ and accumulated color $\mathbf{C}_\text{group}$.
\begin{equation}
    \label{eqn:cached_group}
    T_\text{group} = \prod_{i=1}^{k} (1-\alpha_i),\ \mathbf{C}_\text{group} = \sum_{i=1}^{k}\alpha_i\textbf{c}_i\prod_{j=1}^{i-1}(1-\alpha_j).
\end{equation}
These cached results update $\mathbf{S}(\mathbf{p})$ and $\Gamma$ at group granularity, avoiding recomputing per-Gaussian contributions.

In this way, our algorithm enables a single forward cumulative cache, \cache, to accelerate both forward and backward rasterization while preserving gradient correctness.

\section{\proj Architecture}
\label{sec:arch}

With \cache, we can eliminate a large portion of redundant computations in both forward and backward rasterization.
However, implementing its group identification and cache policy in off-the-shelf GPU software introduces additional shared-memory accesses and control overhead, which can offset the benefits of reuse.
To address this limitation, we propose our hardware architecture, \proj, to efficiently support Gaussian freezing and cached rasterization.
In this section, we first describe the overall architecture.
We then describe the designs of two key components in the \proj core: the rasterization unit array and \cache.

\para{Overview.}
\Fig{fig:arch} shows the overall hardware design of \proj.
We implement \proj as a streaming architecture consisting of five major components: an \proj core, a projection unit, a sorting unit, an aggregation unit, and a postprocessing unit.
The \proj core is our key contribution, which will be further explained in the following paragraphs.
The projection and sorting units execute the projection and sorting stages in the forward pass, similar to prior work~\cite{lee2024gscore, lin2025metasapiens}.
The aggregation unit collects partial gradients for each Gaussian and writes the accumulated results back to memory, like GSArch~\cite{he2025gsarch}.
The postprocessing unit updates parameters and executes densification or freezing according to gradient statistics.


To rank Gaussians by gradients, we reuse the existing sorting unit, so \proj does not require an additional dedicated hardware unit for Gaussian selection.
\proj further adopts tiled SSIM, like~\cite{wen2026cambricon}, so that the image loss can be computed in a streaming manner.
This enables our architecture to pipeline sorting, forward rasterization, and backward rasterization.
To do this, we implement double buffering between consecutive stages, allowing overlap between computation and data communication.

\begin{figure}
    \centering
    \includegraphics[width=\columnwidth]{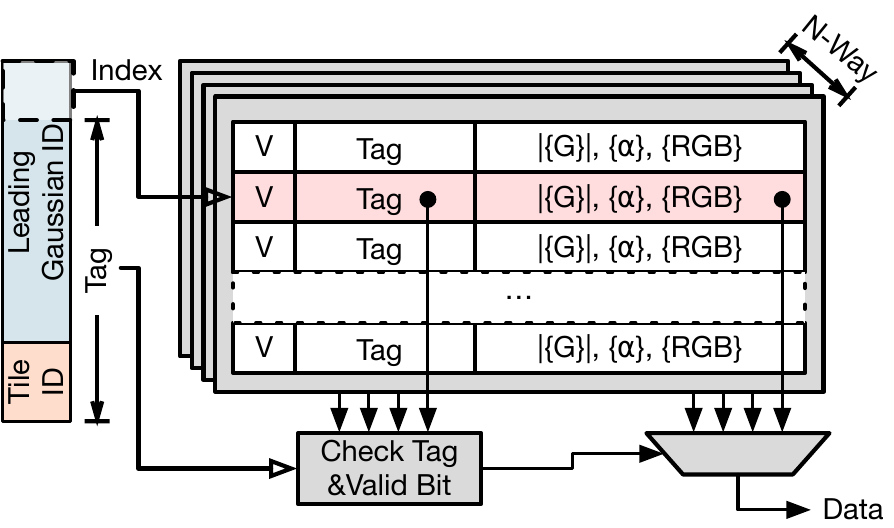}
    \caption{Our \cache design, which is a $N$-way set-associative cache to store frozen Gaussian attributes. Each cache entry stores the length of the consecutive frozen Gaussians and their cumulative transmittance and colors.}
    \Description{The leading Gaussian ID contributes to cache indexing and tagging. Each way has a valid bit, a tag, a group length, and accumulated transmittance and color fields.}
    \label{fig:cache_design}
\end{figure}

\para{\proj Core.}
The key component of our architecture is the \textit{ElsaGS core}, which supports training with our technique, Gaussian freezing, while accelerating both forward and backward rasterization.
As shown in \Fig{fig:arch}, the \proj core consists of five components: a sorted Gaussian buffer, an output buffer, a \cache, a rasterization unit array, and a reduction unit.
Both the Gaussian buffer and the output buffer are double-buffered to support our streaming processing and hide off-chip data traffic latency.
The rasterization unit array performs both forward and backward rasterization computations, while the on-chip reduction unit aggregates partial Gaussian gradients to reduce off-chip memory traffic.

\para{Rasterization Array.}
Our rasterization array comprises a set of rasterization units executing in parallel.
Each rasterization unit adopts a pixel-wise processing paradigm to execute forward or backward rasterization.
The right side of \Fig{fig:arch} shows a single-unit design.
It is organized into two separate stages: a frontend that accumulates Gaussian contributions (light blue) and a backend that computes the image loss or gradients (dark blue).
Each array maintains a temporary buffer, which fetches data from the sorted Gaussian buffer and detects consecutive frozen Gaussians for \cache queries.
This buffer is set to twice the maximum Gaussian-group length used for \cache queries, ensuring sufficient capacity for grouping consecutive frozen Gaussians.

In forward rasterization, the frontend continuously consumes consecutive frozen Gaussians or individual active Gaussians at the top of the temporary buffer, then pops the processed entries and refills the buffer from the sorted Gaussian list.
When the frozen Gaussian group reaches the maximum length, encounters an unfrozen Gaussian, or reaches the end of the list, it queries the \cache.
On a cache hit, the returned cached contribution is directly accumulated into the rendered pixel.
Otherwise, the frontend performs recomputation and updates \cache based on the policy in \Sect{sec:framework:cache}.
Meanwhile, transitional or active Gaussians still go through the canonical forward rasterization pipeline and accumulate their contributions to the pixel.
After the frontend completes the color accumulation over all Gaussians in the sorted Gaussian list, the backend then calculates the image loss and saves it in the buffer.

During backward rasterization, the backend traverses the list in the same order and reconstructs the cumulative color and transmittance.
The backend uses these values and losses in the image buffer to compute partial gradients for active Gaussians.
Our architecture adopts the redesigned backward rasterization in \Sect{sec:framework:raster}, so that the \cache can be shared in both rasterization stages.
Lastly, the partial gradients from all rasterization units are locally aggregated by an on-chip gradient reduction unit to reduce off-chip traffic.

\para{\cache.}
\Fig{fig:cache_design} shows \cache design, which is implemented as a $N$-way associative cache to store precomputed frozen-Gaussian attributes.
For simplicity, we show the cache design for a single view, where the key consists of the tile ID and leading Gaussian ID.

During a cache lookup, we first index the target set through the least significant bits of the group key.
The cache then compares its tags in parallel. A valid tag match identifies a candidate entry.
The rasterization array directly reuses the cached transmittance and color contributions if the stored group length $|\{G\}|$ of the entry matches the current group length.
Length mismatches are handled by the cache update policy.
This way, we can reduce the computation of these consecutive frozen Gaussians.

To further regularize off-chip data access, we design \mbox{\cache} with double buffering.
Each \mbox{\cache} buffer is shared across the rendering of eight image tiles, where each tile has a size of 16$\times$16.
While the current eight tiles are being rendered, the cache contents for the next eight tiles are prefetched into the fill buffer.
In this way, off-chip data movement for \mbox{\cache} becomes streaming.

\section{Experimental Setup}
\label{sec:exp}

\para{Hardware Setup.} 
Our overall architecture consists of 4 projection units, 4 sorting units, 4 gradient reduction units, and a \proj core.
We implement \proj core with 2 rasterization arrays clocked at 1~GHz.
Each has $8\times8$ rasterization units.
Both the sorted Gaussian buffer and the output buffer are shared across all rasterization arrays.
They have a total size of 64~KB and 4~KB, respectively.

The \cache is shared across the rasterization arrays. 
This 512~KB cache has 32 sets with 8 entries per set.
The tag of each cache entry is constructed by the leading Gaussian ID, tile ID, and Image ID.
Each entry stores a tile ($16 \times 16$) of the accumulated color and transparency values.
Each cache access requires 8 cycles to return 512~B of cached data (an $8 \times 8$ tile accumulation result).
Two $8 \times 8$ tiles can be accessed in parallel.
\cache is shared across $8$ image tiles with a size of $16\times16$.
We cap each frozen group at 64 Gaussians. Each rasterization array has a temporary buffer that holds up to 128 Gaussians, twice the maximum group size.

\para{Experimental Methodology.} 
The datapath of \proj is developed using a standard EDA flow with Synopsys and Cadence tools targeting TSMC’s 16~nm FinFET technology.
The resulting design is then scaled to the 8~nm process node using DeepScaleTool~\cite{sarangi2021deepscaletool} to have a fair comparison with the mobile GPU on Nvidia Orin SoC~\cite{stillmaker2017scaling, sarangi2021deepscaletool}.
SRAM components are generated using the Arm Artisan memory compiler, and power estimates are obtained with Synopsys PrimeTimePX using annotated switching activities.
The DRAM subsystem in our simulation is modeled using Micron’s 16~Gb LPDDR5 at 6400~MT/s with four memory channels, following the datasheet~\cite{micronlpddr5}.
DRAM energy consumption is estimated using Micron’s System Power Calculator~\cite{microdrampower}.
To evaluate system-level performance, we then develop a cycle-accurate simulator that models component-level latencies and power consumption obtained from post-synthesis RTL measurements. 

\begin{table} 
\caption{Area comparison. The area is scaled into 8~nm.}
\centering{
\resizebox{0.8\columnwidth}{!}{
\renewcommand*{\arraystretch}{1}
\renewcommand*{\tabcolsep}{5pt}
\begin{tabular}{ cccc } 
\toprule[0.15em]
\textbf{\specialcell{Baselines}} & \textbf{GSArch}~\cite{he2025gsarch} &  \textbf{CambriconGS}~\cite{wen2026cambricon} & \textbf{\proj} \\ 
\midrule[0.05em]
PE (mm$^2$) & 1.04 & 1.57 & 0.96 \\
SRAM (mm$^2$) & 0.11 & 0.13 & 0.38 \\
Total (mm$^2$) & 1.15 & 1.70 & 1.34\\
\bottomrule[0.15em]
\end{tabular}
}
}
\label{tab:area}
\end{table}

\para{Area.}
\Tbl{tab:area} shows that \proj has comparable area compared to the prior 3DGS accelerators~\cite{he2025gsarch, wen2026cambricon}.
The area numbers of GSArch and CambriconGS are obtained from their papers and scaled to 8~nm.
The total area of our design is 1.34~mm$^2$ in 8~nm technology.

\para{Datasets and Metrics.}
We evaluate \proj on three widely adopted real-world datasets: Mip-NeRF360~\cite{barron2022mipnerf360}, Tanks and Temples~\cite{Knapitsch2017}, and DeepBlending~\cite{hedman2018deep}.
Rendering quality is evaluated using standard metrics, including PSNR and SSIM.

\para{Baselines.}
We evaluate \proj using the canonical 3DGS training algorithm~\cite{kerbl20233d}, and a variant with software implementation of atomic operation reduction~\cite{durvasula2025arc}.
For hardware comparison, we consider three baselines. 
The first is \mode{GPU}, a mobile Ampere GPU in Nvidia’s Orin SoC. 
The other two are \mode{GSArch}~\cite{he2025gsarch} and \mode{CamGS}~\cite{wen2026cambricon}, both of which are dedicated accelerators designed for 3DGS training.
We do not compare against inference-only accelerators~\cite{lee2024gscore, pei2025gcc, ye2025gaussian, feng2025lumina, li2026orange}, as they do not support the training process.

\para{Variants.}
To dissect our contributions in algorithm and architecture, we evaluate two variants of \proj to separate the contributions proposed in our paper. We set the Gaussian freezing ratio $\rho = 90\%$ and the finetuning stage to start at $I_{th}=15k$:
\begin{itemize}
\item \mode{\proj-sw}: GPU implementation of Gaussian freezing.
\item \mode{\proj-hw}: this variant is a full-fledged implementation with our hardware support.
\end{itemize}

\begin{figure}
    \centering
    \includegraphics[width=\columnwidth]{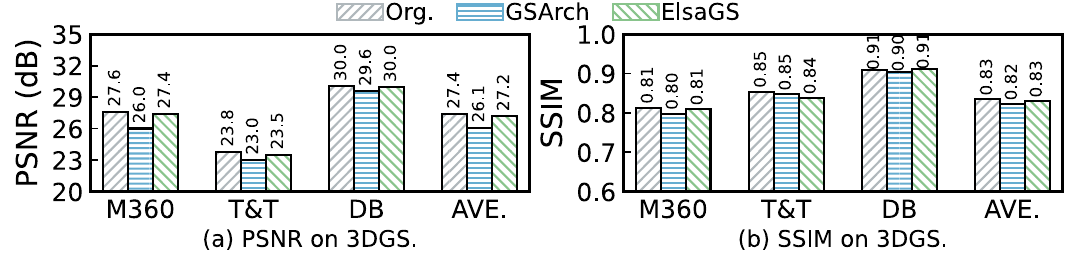}
    \caption{Quality evaluation over 3DGS.}
    \Description{The two panels compare PSNR and SSIM for the original pipeline, GSArch, and ElsaGS across three datasets.}
    \label{fig:acc}
\end{figure}

\begin{figure}
    \centering
    \includegraphics[width=\columnwidth]{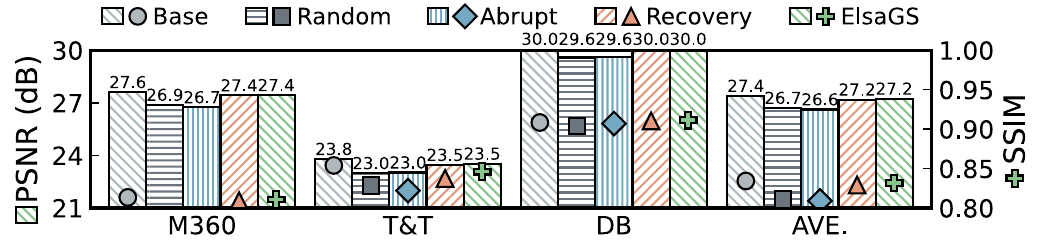}
    \caption{Freezing framework ablation. \mode{Base}: canonical 3DGS. \mode{Random}: freeze Gaussians randomly under the progressive freezing budget. \mode{Abrupt}: freeze $\rho$ of Gaussians at every freezing event. \mode{Recovery}: unfreeze all Gaussians and finetune for 1k iterations before final freezing step.}
    \Description{Ablation study of freezing framework.}
    \label{fig:abl_acc}
\end{figure}

\section{Evaluation}
\label{sec:eval}

\subsection{Quality Evaluation}
\label{sec:eval:acc}

\Fig{fig:acc} compares the reconstruction quality of \proj against two baselines: the original 3DGS pipelines and the technique proposed by \mode{GSArch}.
Here, we evaluate them on canonical 3DGS.
Overall, \mode{\proj} achieves nearly identical performance to the original 3DGS. 
\proj achieves a PSNR 0.2~dB below canonical 3DGS, 1.08~dB above GSArch.
For SSIM, \proj shows a similar result, while \mode{GSArch} loses 0.013.
\Fig{fig:abl_acc} compares different freezing strategies (see \Fig{fig:abl_acc} caption). 
Compared to \proj, \mode{Random} and \mode{Abrupt} yield 0.5~dB and 0.6~dB lower PSNR, respectively.
Thus, our training strategy can better preserve quality compared to other heuristics.
On the other hand, \mode{Recovery} provides negligible improvement over \mode{\proj}, supporting our choice to keep selected Gaussians frozen under the training schedule.

\subsection{Performance}
\label{sec:eval:perf}

\begin{figure}[t]
\centering
\includegraphics[width=0.98\columnwidth]{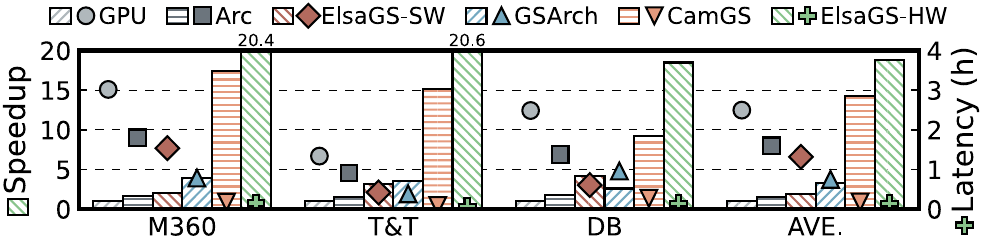}
\caption{End-to-end training latency. Speedup numbers are normalized to the corresponding \mode{GPU} baselines.}
\Description{Bars and markers compare six methods across three datasets and an aggregate group.}
\label{fig:speedup}
\end{figure}

\para{Speedup.}
\Fig{fig:speedup} compares end-to-end training speed across hardware platforms.
Overall, \mode{\proj-hw} achieves the highest speedup across all datasets and training algorithms.
On average, \mode{\proj-hw} delivers 18.7$\times$ speedup over \mode{GPU}, reducing the training time from hours to minutes.
Compared to dedicated accelerators, \mode{\proj-hw} achieves 5.6$\times$ and 1.3$\times$ speedup over \mode{GSArch} and \mode{CamGS}, respectively.
Meanwhile, we also report the speedup of our GPU-based implementation, \mode{\proj-sw}, which achieves 1.9$\times$ and 1.6$\times$ speedup over the \mode{GPU} and \mode{Arc}, respectively, by omitting gradient computation and accumulation of frozen Gaussians.

\begin{figure}[t]
    \centering
    \includegraphics[width=\columnwidth]{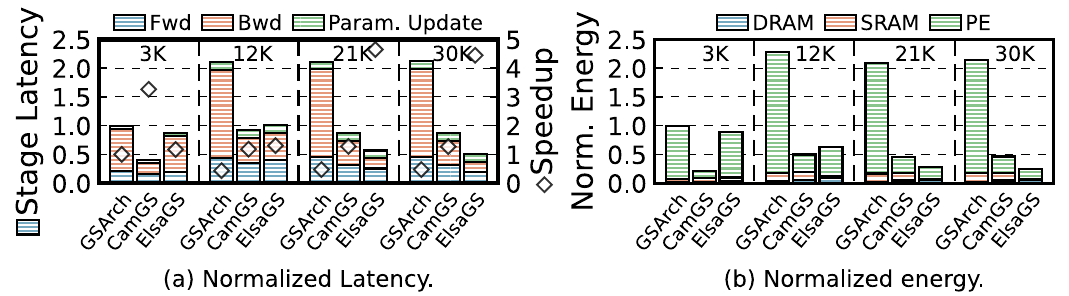}
    \caption{Normalized latency, per-frame speedup and energy over training iterations (3k, 12k, 21k, 30k). All numbers are normalized to \mode{GSArch} at iteration 3k.}
    \Description{\zhuhe{The left panel shows stage latency bars and frame speedup markers. The right panel shows dynamic energy bars at 3K, 12K, 21K, and 30K iterations.}}
    \label{fig:eval_details}
\end{figure}

\Fig{fig:eval_details}a compares stage latency and per-frame speedup over iterations. 
At initial training, all Gaussians are active, and \mode{\proj-hw} is slower than \mode{CamGS}. 
As training progresses, more Gaussians are frozen, and their contributions are cached in \cache.
\mode{\proj-hw} reduces computation in both forward and backward rasterization. 
E.g., at iteration 12k, \mode{\proj-hw} achieves per-frame speedups of 1.1$\times$ and 3.0$\times$ over \mode{CamGS} and \mode{GSArch}, respectively.
At fine-tuning, the freezing ratio reaches 90\%.
Compared to \mode{GSArch} and \mode{CamGS}, \mode{\proj-hw} achieves at least 9.4$\times$ and 3.5$\times$ speedup.


\begin{figure}[t]
\centering
\includegraphics[width=0.98\columnwidth]{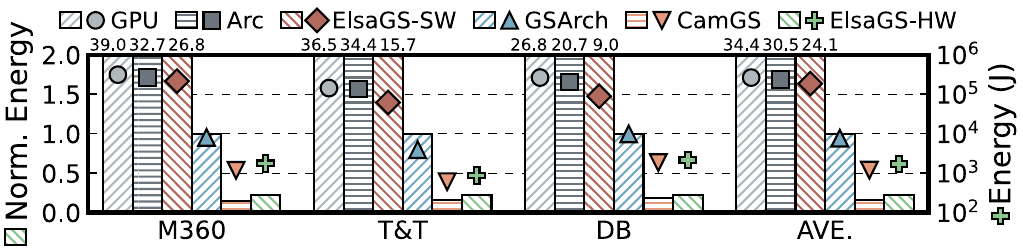}
\caption{End-to-end training energy. Bars show energy normalized to \mode{GSArch}.}
\Description{\zhuhe{Bars and markers compare the training energy of six methods across three datasets and an aggregate group.}}
\label{fig:energy}
\end{figure}

\para{Energy Savings.}
\Fig{fig:energy} shows the end-to-end training energy of different hardware across datasets and training methods.
Overall, \mode{\proj} reduces the training energy by 78.0\% compared to \mode{GSArch} and achieves comparable energy consumption to \mode{CamGS}.
Compared to \mode{GPU}, \mode{\proj} reduces the training energy by 99.4\% on average.
\Fig{fig:eval_details}b further shows the normalized per-frame energy over iterations. 
As the training progresses, \mode{\proj-hw} gains a per-frame energy advantage over other accelerators.
Since \cache reduces a large amount of computation in both forward and backward passes, \mode{\proj-hw} consumes over 40\% less dynamic energy per frame than \mode{CamGS} and over 85\% less than \mode{GSArch} at finetuning stage.



\begin{figure}[t]
\centering
\begin{minipage}[t]{0.48\columnwidth}
    \centering
    \includegraphics[width=\columnwidth]{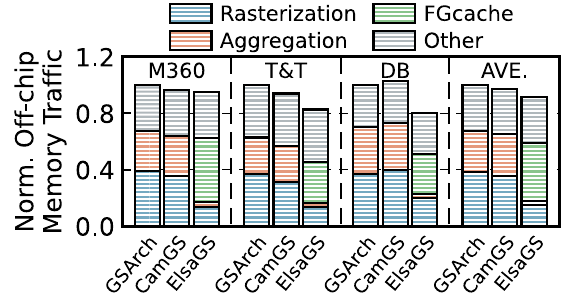}
    \caption{Off-chip Memory Traffic. Numbers are normalized to \mode{GSArch}.}
    \Description{A dual-axis plot shows speedup and normalized energy for groups of 2, 4, 8, and 16 tiles.}
    \label{fig:mem_traffic}
\end{minipage}
\hfill
\begin{minipage}[t]{0.48\columnwidth}
    \centering
    \includegraphics[width=\columnwidth]{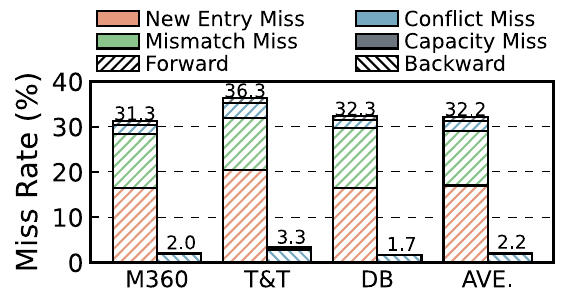}
    \caption{Study of \cache miss rate on forward and backward rasterization.}
    \Description{Stacked bars separate forward and backward misses into new-entry, mismatch, conflict, and capacity categories.}
    \label{fig:miss_rate}
\end{minipage}
\end{figure}

\para{Memory Traffic.}
In \Fig{fig:mem_traffic}, we report the off-chip memory traffic of \proj under a 90\% freezing ratio.
We break down the traffic into four categories.
The first category is the memory traffic in forward and backward rasterization.
The second one is the data transmission during aggregation, where the pipeline accumulates gradients through atomic operations.
The third one is the data transmission caused by the \cache, which only occurs in \proj.
The last one includes the remaining memory traffic.

As the results show, \proj reduces most memory traffic in the aggregation stage.
This is because \proj leverages Gaussian-wise sparsity compared to other methods, naturally reducing most gradient accumulation.
Moreover, \proj reduces over 50\% memory traffic in rasterization by caching the cumulative contributions. 
Meanwhile, our overall cache design effectively reduces the cache traffic and further reduces the overall memory traffic.

\para{Cache Miss Rate.}
\Fig{fig:miss_rate} gives a case study of cache miss rates on 3DGS with 90\% Gaussians frozen.
We break down misses by forward and backward rasterization and further classify these misses into four categories.
The first one is the conflict misses due to new cache entries.
This comes from the original grouped Gaussians being split into two.
The second one is the conflict misses due to cache entry mismatches. 
This is from the grouped Gaussians having been expanded or shortened.
The third one is the canonical conflict misses.
These are misses similar to the conventional cache conflict misses due to cache associativity.
The last one is capacity misses.

Most misses occur in the forward pass.
Among them, 64.8\% are classified as conflict misses.
However, most of these conflicts come from the first two categories, i.e., new entries and entry mismatches, rather than conventional cache conflicts.
These misses are mainly caused by the dynamic nature of 3DGS training.
As Gaussian parameters are updated over iterations, active Gaussians can alter the ordering of existing Gaussian groups.
In contrast, entry-mismatch and new-entry misses do not appear in the backward pass, because Gaussian parameters remain unchanged in the same iteration and backward rasterization can fully reuse cached results.

\subsection{Ablation Study}
\label{sec:eval:abl}

\begin{figure}[t]
\centering
\begin{minipage}[t]{0.48\columnwidth}
    \centering
    \includegraphics[width=\columnwidth]{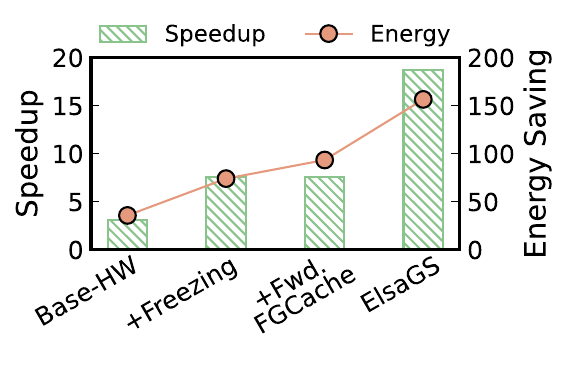}
    \caption{Ablation study with different variants. Numbers are normalized to \mode{gpu}.}
    \label{fig:hw_ablation}
\end{minipage}
\hfill
\begin{minipage}[t]{0.48\columnwidth}
    \centering
    \includegraphics[width=\columnwidth]{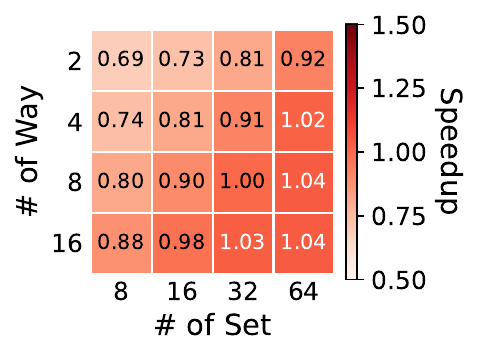}
    \caption{Sensitivity of performance to various cache configurations.}
    \label{fig:cache_config}
\end{minipage}
\end{figure}

\Fig{fig:hw_ablation} further dissects the contribution of \mode{\proj-hw}.
Specifically, we gradually include more components in \proj, \mode{+Freezing} and \mode{+Fwd. \cache}.
\mbox{\mode{Base-hw}} runs with all Gaussians active and \cache disabled, which improves performance over \mbox{\mode{GPU}} by 3.1$\times$.
\mode{+Freezing} skips gradient computation and accumulation for frozen Gaussians, yielding a 2.4$\times$ speedup and reducing energy by 51.9\% relative to \mode{Base-hw}.
\mode{+Fwd. \cache} only supports \cache for forward rasterization and further reduces energy by 20.8\% relative to \mode{+Freezing}, with negligible change in speedup because backward rasterization still recomputes the contributions of frozen Gaussians.
Finally, \mode{\proj} adds the cache-compatible backward rasterization in Sec. 4.3, achieving a further 2.5$\times$ speedup and reducing energy by 40.3\% relative to \mode{+Fwd. \cache}.

\begin{figure}[t]
\centering
\begin{minipage}[t]{0.48\columnwidth}
    \centering
    \includegraphics[width=\columnwidth]{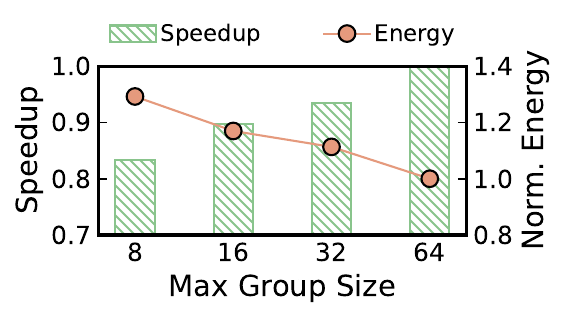}
    \caption{Sensitivity of speedup and energy consumption to the maximum frozen group size.}
    \Description{A dual-axis plot compares speedup and energy for maximum group sizes of 8, 16, 32, and 64 Gaussians.}
    \label{fig:sensitivity_max_groupsize}
\end{minipage}
\hfill
\begin{minipage}[t]{0.48\columnwidth}
    \centering
    \includegraphics[width=\columnwidth]{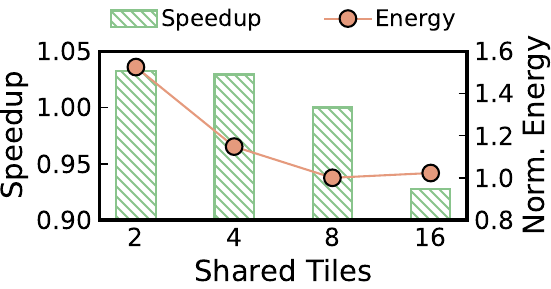}
    \caption{Performance Sensitivity to number of shared tiles in \cache. Numbers are normalized to the 8-tile configuration.}
    \Description{A dual-axis plot shows speedup and normalized energy for groups of 2, 4, 8, and 16 tiles.}
    \label{fig:sensitivity_tile_sharing}
\end{minipage}
\end{figure}

\subsection{Sensitivity Study}
\label{sec:eval:sens}

\para{Shared Tiles per \cache.}
\Fig{fig:sensitivity_tile_sharing} shows the sensitivity of performance to the number of shared tiles per \cache, normalized to the 8-tile configuration.
As the number of shared tiles increases, overall speedup decreases due to higher cache conflict rates.
Meanwhile, energy efficiency is also improved as tile sharing increases from 2 to 8.
However, further increasing the number of shared tiles from 8 to 16 decreases both speedup and energy efficiency due to increasing cache contention.


\para{Cache Configuration.}
\Fig{fig:cache_config} shows the sensitivity of speedup to cache configuration with 90\% of Gaussians frozen.
All numbers are normalized to the configuration with 32 sets and 8 ways.
The result shows that increasing either the number of sets or associativity generally improves speedup by reducing cache conflicts.
However, the overall performance gains diminish beyond our default configuration, i.e., 32 sets and 8 ways, where further increases yield marginal gains.
This shows that our default configuration achieves a good balance between performance and hardware cost.

\begin{figure}[t]
\centering
\begin{minipage}[t]{0.48\columnwidth}
    \centering
    \includegraphics[width=\columnwidth]{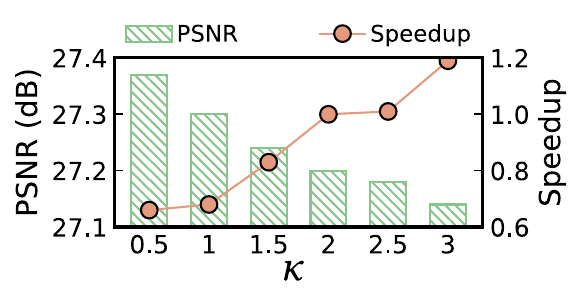}
    \caption{Sensitivity to the freezing multiplier, $\kappa$.}
    \Description{A dual-axis plot shows peak signal-to-noise ratio and speedup as the freezing multiplier varies from 0.5 to 3.}
    \label{fig:sensitivity_multiplier}
\end{minipage}
\hfill
\begin{minipage}[t]{0.48\columnwidth}
    \centering
    \includegraphics[width=\columnwidth]{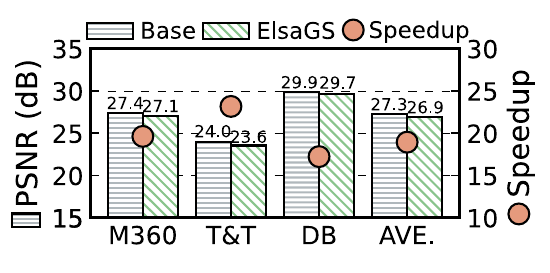}
    \caption{Evaluation of \mode{\proj} on Taming3DGS.
    }
    \Description{A scatter plot places reconstruction quality on the horizontal axis and speedup on the vertical axis. Markers are labeled with the pruning or freezing ratio.}
    \label{fig:perf_taming}
\end{minipage}
\end{figure}

\para{Freezing Multiplier.}
\Fig{fig:sensitivity_multiplier} shows how the multiplier $\kappa$ in \Eqn{eqn:frozen} affects speed and rendering quality.
We fix the freezing cap $\rho$ at 90\%.
Speedup is normalized to the default configuration with $\kappa = 2$.
As $\kappa$ increases, \proj freezes Gaussians more aggressively during training, leading to higher speedup.
Meanwhile, aggressively freezing Gaussians slightly reduces reconstruction quality.

\para{Maximum Frozen Group Size.}
\Fig{fig:sensitivity_max_groupsize} shows the effect of the maximum frozen group size on speedup and energy consumption, normalized to the configuration with a maximum group size of 64 Gaussians.
Here, we keep the default cache configuration.
As the maximum group size increases, speedup gradually increases because grouping more consecutive frozen Gaussians can reduce cache contention.
At the same time, normalized energy decreases due to reduced recomputation.

\begin{figure}[t]
\centering
\begin{minipage}[t]{0.48\columnwidth}
    \centering
    \includegraphics[width=\columnwidth]{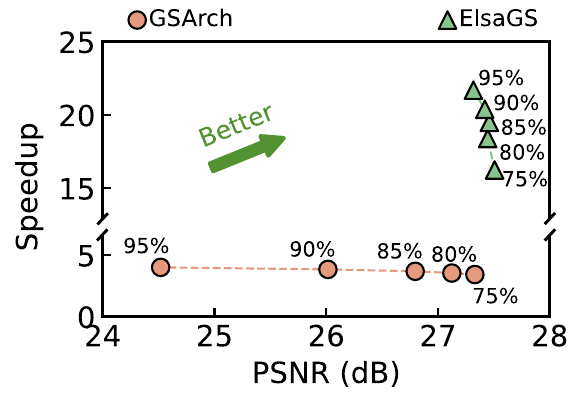}
    \caption{Comparison against \mode{GSarch} across different pruning/frozen ratios. 
    }
    \Description{A scatter plot places reconstruction quality on the horizontal axis and speedup on the vertical axis. Markers are labeled with the pruning or freezing ratio.}
    \label{fig:gsarch_compare}
\end{minipage}
\hfill
\begin{minipage}[t]{0.48\columnwidth}
    \centering
    \includegraphics[width=\columnwidth]{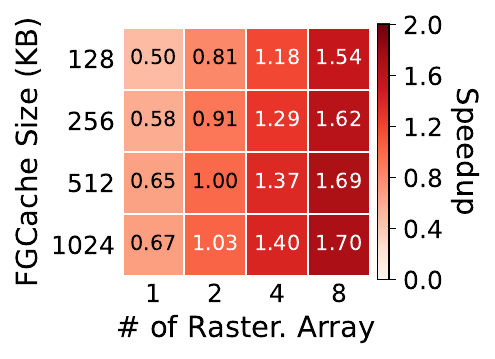}
    \caption{Scalability analysis on \cache size and number of rasterization arrays.}
    \Description{A heatmap gives speedup for combinations of 128 to 1,024 kilobytes of cache and one to eight rasterization arrays.}
    \label{fig:scalability}
\end{minipage}
\end{figure}

\subsection{Scalability}
\label{sec:eval:scale}

\Fig{fig:scalability} shows the scalability of \proj to \cache size and the number of rasterization arrays under Taming3DGS.
Here, we set the cache associativity to be 8 across all cache sizes.
The speedup numbers are normalized to our default configuration, 2 rasterization arrays with a cache of 512~KB.
It shows that increasing the number of rasterization arrays improves speedup.
However, excessive rasterization arrays would lead to sublinear scaling due to the bottleneck shifts.
Meanwhile, larger \cache sizes could further improve performance by alleviating cache contention and reducing recomputation, with the size of 512~KB being a good tradeoff.

\subsection{Performance on Taming3DGS}
\label{sec:eval:taming}

We further evaluate the performance of \proj under the same experimental settings on Taming3DGS~\cite{mallick2024taming}, a points pruning variant of 3DGS. 
Compared to the baseline, \proj achieves 19.0$\times$ speedup with 0.4~dB PSNR loss on average.
The performance gain suggests that \proj can be applied to other 3DGS-based algorithms by leveraging similar freezing and caching techniques.


\subsection{Comparison with GSArch}
\label{sec:eval:gsarch}


\Fig{fig:gsarch_compare} shows how freezing and pruning affect 3DGS training. The evaluation is performed on M360 dataset.
From 75\% to 95\%, \proj is substantially faster than GSArch at comparable or higher PSNR.
Increasing the freezing ratio $\rho$ improves speedup with only a modest loss in reconstruction quality.
In contrast, \mode{GSArch} achieves limited speedup with noticeable PSNR drops at high pruning ratios.


\section{Related Work}

\para{Neural Rendering Accelerators.}
In recent years, a wide range of studies have focused on accelerating neural rendering~\cite{lee2023neurex, rao2022icarus, li2023instant, mubarik2023hardware, li2022rt, fu2023gen, feng2024cicero, song2024srender, liu2025cambricon, feng2025lumina, ye2025gaussian, lee2024gscore, li2025uni, lee2025vr, lin2025metasapiens, durvasula2025arc, he2025gsarch, li2023sltarch, zhang2025streaming, wen2026cambricon, pei2025gcc, zhao2026clm, lee2026gs}.
Some studies~\cite{lee2025vr, feng2025lumina, lin2025metasapiens, zhu2026nebula, oh2026neo, liu2026asdr} propose dedicated accelerators for rendering only.
Meanwhile, others~\cite{he2025gsarch, wu2024gauspu, huang2026splatonic} focus on optimizing the training procedure.
REACT3D~\cite{wang2025react}, RTGS~\cite{li2025rtgs}, and Splatonic~\cite{huang2026splatonic}, propose dedicated accelerators for SLAM-based 3DGS.

\para{3DGS Acceleration Techniques.}
In addition to hardware support, numerous software optimizations have been proposed to accelerate 3DGS.
To reduce rendering workload, studies~\mbox{\cite{fang2024mini, fang2024mini2, mallick2024taming, fan2023lightgaussian, hanson2025pup}} propose to prune insignificant Gaussians.
Others~\mbox{\cite{lee2024compact, niedermayr2024compressed, niemeyer2024radsplat, ren2024octree, kerbl2024hierarchical}} introduce alternative data representations to alleviate memory pressure.
Another line of research targets optimizing the 3DGS pipeline itself, e.g., online filtering techniques~\cite{klosowski1998efficient, wang2024adr, hanson2025speedy, gottschalk1996obbtree, lee2024gscore} or rasterization optimization~\cite{feng2025flashgs, liao2025tc, gui2024balanced, zhu2026seele}.
Recently, GSta~\cite{armagan2026gsta} and SparseOIT~\cite{yang2026sparseoit} attempt to accelerate training by fixing the parameters of a subset of Gaussians. 
However, they either only optimize the backward pass or rely on order-independent transparency rendering assumptions. 
In contrast, \proj exploits the unique training characteristics to accelerate both forward and backward passes, and is applicable to different 3DGS-based algorithms.


\para{Memoization Techniques.}
There is a long history in the architecture community of leveraging memoization to reduce computation.
Recent work~\cite{hegde2018ucnn, riera2018computation, azizimazreah2019shortcut, brock2017freezeout, wang2023egeria} has explored computation reuse for vision DNNs at multiple granularities.
Other works~\cite{liu2023space, liu2025generalizing} target resource-constrained platforms by clustering similar inputs to avoid redundant convolution operations.
Meanwhile, another line of research exploits temporal correlations in video streams to bypass unnecessary computation~\cite{zhu2018euphrates, ying2022exploiting, feng2019asv, zhao2020deja, zhao2021holoar, ma2024deepcache}.

\section{Conclusion}
\label{sec:conc}

We propose \proj, accelerates 3DGS training by progressively freezing low-gradient Gaussians and reusing their cached contributions with negligible quality loss.
To improve the reuse efficiency, we develop \cache and cache-compatible backward rasterization.
We further introduce a streaming architecture, \proj core, with our dedicated frozen-Gaussian cache.
Our results show that \proj freezes up to 90\% of Gaussians with higher speedup against off-the-shelf GPUs and prior accelerators.

\bibliographystyle{ACM-Reference-Format}
\bibliography{refs}

\end{document}